\documentclass[%
 pra,
longbibliography,
amsmath,amssymb,
 reprint,%
]{revtex4-2}

\usepackage{graphicx}
\usepackage{xcolor}
\usepackage{dcolumn}
\usepackage{bm}

\usepackage[utf8]{inputenc}
\usepackage[T1]{fontenc}
\usepackage{mathptmx}
\usepackage{etoolbox}
\usepackage[version=4]{mhchem}
\usepackage{stfloats}
\usepackage{graphicx}
\usepackage{subcaption}
\usepackage{float}
\usepackage{placeins}
\usepackage[justification=centering,singlelinecheck=false]{caption}

\makeatletter
\def\@email#1#2{%
 \endgroup
 \patchcmd{\titleblock@produce}
  {\frontmatter@RRAPformat}
  {\frontmatter@RRAPformat{\produce@RRAP{*#1\href{mailto:#2}{#2}}}\frontmatter@RRAPformat}
  {}{}
}%
\makeatother
\begin{document}

\preprint{AIP/123-QED}

\title{Size Effects in the Strong-Field Ionization and Dissociation Dynamics of (H$_2$O)$_n$ (n=1–4)}
\author{Chen Jiang}
\affiliation{Department of Physics and Astronomy, Vanderbilt
University, Nashville,
Tennessee 37235, United States}
\author{Cody L. Covington}
\affiliation{ Department of Chemistry, Austin Peay State University,
Clarksville,
Tennessee 37044, United States}
\author{Kalman Varga}
\email{kalman.varga@vanderbilt.edu (corresponding author)}
\affiliation{Department of Physics and Astronomy, Vanderbilt
University, Nashville,
Tennessee 37235, United States}

\date{\today}

\begin{abstract}
The size-dependent strong-field ionization and dissociation dynamics of 
(H$_2$O)$_n$ (n=1–4) are investigated using real-time time-dependent density 
functional theory (RT-TDDFT) coupled to Ehrenfest molecular dynamics under 
a common few-cycle near-infrared laser pulse. It is found that the net 
ionization per monomer varies only weakly on cluster size, whereas 
the protonic and oxygen response is changed much more strongly once the 
cluster size grows beyond the dimer. In particular, H-ejection activity is 
observed to rise sharply from the dimer to the trimer/tetramer regime, 
while stable H-transfer is essentially absent in the dimer under the 
present criterion but becomes substantial in the trimer and is further 
amplified in the tetramer. Through timing analyses, it is shown that the 
dimer exhibits a weak and temporally broad response, whereas the larger 
clusters display a much stronger early-time protonic response concentrated 
within and immediately after the laser pulse window. By endpoint oxygen 
statistics, a systematic increase in dissociation propensity with cluster 
size is likewise shown. For a clean subset of direct two-body dimer breakup 
trajectories, the asymptotic kinetic energy release is estimated to be
4.47 $\pm$ 1.03 eV, in reasonably good agreement with the experimental 
value for the unprotonated two-body Coulomb-explosion channel. Overall, 
it is shown by the results that increasing water-cluster size primarily 
reshapes the strong-field response through proton-mediated and topology-level nuclear
dynamics rather than through a large change in net ionization alone.
\end{abstract}

\maketitle

\section{\label{sec:level1}Introduction}

Hydrogen-bond networks are ubiquitous in chemistry and biology, and their response to ionization
governs processes ranging from early-stage radiation damage to proton-transfer signaling in
biomolecular environments. In biological systems, the disruption and reorganization of
hydrogen-bond networks following ionization can influence radiation-induced strand breaks,
intramolecular proton-transfer pathways, and the structural integrity and reactivity of
macromolecules.\cite{Hahn2021CommunChem,Reisz2013RadiatRes,Kratochvil2023NatChem}
These connections have spurred sustained interest in understanding the ultrafast dynamics of
hydrogen-bonded systems after ionization, and a rich body of experimental and theoretical work
has emerged in recent years.\cite{Mucke2010,Jahnke2015,Cederbaum2003,Thuermer2013,
PhysRevA.98.050701,Zhang2025,PhysRevLett.128.133001}

Among hydrogen-bonded systems, small water clusters have served as prototypical models for
isolating the elementary steps of ionization-driven dynamics in a controlled setting. A
central theme in this literature is ultrafast proton transfer (PT), which has been studied
extensively in clusters ranging from the dimer to larger
oligomers.\cite{Zhang2019PRA,Tachikawa2011PCCP,Schnorr2023SciAdv,Wang2025JCP,Barnett1995JPC,Radi1999JCP,
Svoboda2013PCCP,Chalabala2018JPCA,Tachikawa2015RSCAdv,Wang2014LaserPhys,Zhang2024PRA_NH3,
2t5k-6n9s}
The water dimer has been established as a particularly well-characterized benchmark for
strong-field studies of hydrogen bonding.\cite{Zhang2019PRA} In a landmark experiment,
Zhang \textit{et al.} irradiated isolated water dimers with 780-nm, $38 \pm 2$~fs laser pulses
at a peak intensity of $(1.2 \pm 0.2)\times10^{14}$~W/cm$^2$ and resolved the
proton-transfer dynamics through the branching ratio between the protonated and unprotonated
two-body Coulomb-explosion channels, obtaining a proton-transfer time constant of
$31 \pm 5$~fs for the singly charged dimer ion.\cite{Zhang2019PRA}

{\color{black}More recently, Schnorr \textit{et al.} employed XUV pump--probe ion coincidence spectroscopy to directly track proton transfer in isolated water dimers following single-photon ionization, measuring a state-averaged proton-transfer time of $55\pm20$~fs and providing an important benchmark for ultrafast proton dynamics in the water dimer~\cite{Schnorr2023SciAdv}. Their result complements the strong-field dimer measurements of Zhang \textit{et al.} by probing the same elementary proton-transfer process in a perturbative ionization regime.}

Parallel theoretical work has explored whether the timescale and mechanism of proton transfer
depend systematically on cluster size. Using direct \textit{ab initio} molecular-dynamics
simulations on $(\mathrm{H_2O})_n^+$ ($n = 2$--4), Tachikawa and Takada
\cite{Tachikawa2015RSCAdv} demonstrated a pronounced size dependence in the initial
proton-transfer step, with mean transfer times of 28.4, 15.1, and 9.9~fs for the dimer,
trimer, and tetramer, respectively. They further identified a secondary proton-transfer
event occurring exclusively in the larger clusters, with mean times of 119.6~fs (trimer)
and 39.6~fs (tetramer), suggesting that the extended hydrogen-bond network introduces
qualitatively new protonic pathways absent in the dimer.\cite{Tachikawa2015RSCAdv}

Despite this progress, two important limitations remain. First, most studies have focused
either on the water dimer in isolation or on singly ionized water-cluster cations prepared in
a fixed charge state, leaving the question of how cluster size reshapes the response under
genuine strong-field, multi-ionization conditions largely unresolved. Second, the discussion
has centered predominantly on proton transfer, while other key observables---net ionization,
proton ejection, and oxygen-framework dissociation---and their coupled evolution across cluster
sizes have received comparatively little attention. A unified, size-resolved picture of the
complete strong-field response, from initial electron removal through the ensuing protonic and
structural dynamics, is therefore still missing.

{\color{black}In particular, it remains unclear which aspects of the strong-field response are largely inherited from the single-molecule ionization physics and which emerge collectively from the extended hydrogen-bond network.}

In the present work, we address this gap by investigating the size-dependent strong-field
ionization and dissociation dynamics of $(\mathrm{H_2O})_n$ ($n = 1$--4) under a common
few-cycle near-infrared laser pulse.
{\color{black}This approach complements recent dimer pump--probe measurements by extending the comparison across cluster size within a common strong-field framework.} By applying identical pulse conditions across all cluster
sizes, we isolate the role of hydrogen-bond-network topology and size on the ultrafast response,
examining how net ionization per monomer, proton ejection, proton transfer, and
oxygen-framework dissociation each evolve from the monomer through the tetramer. This
comparative approach reveals which aspects of the response are set primarily by the single-molecule
strong-field physics and which emerge collectively from the extended network.

To treat the coupled electron--nuclear dynamics under explicit field driving without
invoking precomputed potential-energy surfaces, we employ real-time time-dependent density
functional theory \cite{PhysRevLett.52.997,Ullrich2011TDDFTbook} on a real-space
grid, coupled to Ehrenfest molecular dynamics. This framework propagates the Kohn--Sham
orbitals and nuclear positions self-consistently under the time-dependent Hamiltonian,
capturing charge migration, state mixing, and energy flow on attosecond--femtosecond
timescales.\cite{ComNano,Attosecond} It thereby places strong-field ionization, protonic
response, and oxygen dissociation on the same dynamical footing---a prerequisite for the
kind of unified, multi-observable analysis pursued here. The method has previously been
applied to dissociation dynamics in hydrogen-bonded
clusters,\cite{Jiang2026PRA,Wang2025JCP} laser-driven fragmentation across a range of
molecular systems,\cite{Taylor2025PRA1,Taylor2025PRA2,Jiang2025JCP,Russakoff2015PRA,
Russakoff2015PRA2,Bubin2011LaserAssistedDesorption,Bubin2012GrapheneGraphaneLaser}
and ultrafast protonic motion,\cite{Jiang2025JCP,Wang2025JCP,Taylor2025HydrogenGraphene,
Bubin2011LaserAssistedDesorption} establishing its suitability for the present problem.

\section{\label{sec:level2}Computational Method}
To simulate the dissociation dynamics, we performed an ensemble of independent
calculations initialized with randomized atomic velocity distributions
and molecular orientations, propagating the atomic trajectories via
real-time time-dependent density functional theory coupled to Ehrenfest dynamics.
The Kohn–Sham Hamiltonian has the following form:

\begin{equation}
\begin{split}
\hat H_{\mathrm{KS}}(t)
= {}& -\frac{1}{2}\nabla^{2}
     +V_{\mathrm{ion}}(\mathbf r,t)
     +V_{H}[\rho](\mathbf r,t) \\
  & +V_{\mathrm{XC}}[\rho](\mathbf r,t)
     +V_{\mathrm{laser}}(\mathbf r,t),
\end{split}
\tag{1}
\end{equation}
Here, the first term, $-\nabla^{2}/2$ (in atomic units), is the single-electron kinetic-energy
operator.
The electron density is  
\begin{equation}
\rho(\mathbf r,t)=
   \sum_{k=1}^{N_{\text{occ}}}f_k
   \bigl|\psi_{k}(\mathbf r,t)\bigr|^{2},
\tag{2}
\end{equation}
and the sum runs over all $N_{\text{occ}}$ occupied Kohn–Sham orbitals.
$V_{\mathrm{ion}}$ denotes the external potential due to the ions, modeled with
norm-conserving pseudopotentials centered on each nucleus as given by Troullier
and Martins \cite{Troullier1991PRB}.
$V_{H}$ is the Hartree potential, representing the classical electrostatic Coulomb
interaction among electrons,
\begin{equation}
V_{H}(\mathbf r,t)=
  \int \frac{\rho(\mathbf r^{\prime},t)}
            {|\mathbf r-\mathbf r^{\prime}|}\,
        d\mathbf r^{\prime},
\tag{3}
\end{equation}
where atomic units ($e=m_e=\hbar=1$) are used throughout Eqs.~(1)--(10) unless stated otherwise.
The exchange-correlation potential $V_{\mathrm{XC}}$ is
approximated using the generalized gradient approximation (GGA), 
developed by Perdew et al. \cite{Perdew1986PRB,Becke1988PRA}.
Density functionals contain self-interaction error where
electrons spuriously interact with their own charge distribution.
To remove self-interaction, we employ the ADSIC approach
\cite{Legrand_2002}, which subtracts a fraction $1/N$ from the total density. The remaining
density, $\rho(N-1)/N$, then characterizes the density of all other
electrons as seen by a given spectator electron.
{\color{black}This correction is particularly important in strong-field ionization calculations because self-interaction errors can artificially lower ionization potentials and overdelocalize the emitted charge density.}
The last term in Eq.~(1), $V_{\mathrm{laser}}(\mathbf r,t)$, is the time-dependent potential induced by the laser electric field.
Within the dipole approximation, it is written as $V_{\mathrm{laser}}(\mathbf r,t)=\mathbf r\cdot\mathbf E_{\mathrm{laser}}(t)$. The electric field $\mathbf E_{\mathrm{laser}}(t)$ is taken to be 
\begin{equation}
\mathbf E_{\mathrm{laser}}(t)
= E_{\max}\exp\!\left[-\frac{(t-t_{0})^{2}}{2a^{2}}\right]\sin(\omega t)\,\hat{\mathbf{e}},
\tag{4}
\end{equation} 
where $E_{\max}$, $t_{0}$, and $a$ specify the peak amplitude, the temporal center of the pulse, and the $1/e$ half-width of the Gaussian envelope, respectively.
The parameter $\omega = 2\pi c/\lambda$ denotes the angular laser frequency, and $\hat{\mathbf{e}}$ is the unit vector along the linear polarization direction of the electric field.

Before the time-dependent calculations, we perform a
density-functional-theory (DFT) calculation to obtain the ground state of the
system, including the equilibrium electron density, self-consistent Kohn–Sham
orbitals, and the total energy.  With these initial conditions, we propagate the
orbitals in time using the time-dependent Kohn–Sham equation
\begin{equation}
i\,
\frac{\partial\psi_{k}(\mathbf r,t)}{\partial t}
      =\hat H_{\mathrm{KS}}(t)\,\psi_{k}(\mathbf r,t),
\tag{5}
\end{equation}
which is integrated with the propagator
\begin{equation}
\psi_{k}(\mathbf r,t+\delta t)=
   \exp\!\Bigl[-\,i\hat H_{\mathrm{KS}}\!\left(t+\tfrac{\delta t}{2}\right)\delta t\Bigr]
   \psi_{k}(\mathbf r,t),
\tag{6}
\end{equation}
where $\hat{H}_\mathrm{KS}$ is evaluated at the midpoint $t+\delta t/2$ to maintain second-order accuracy in time.
The propagator is approximated by a fourth-order Taylor expansion:
\begin{equation}
\psi_{k}(\mathbf r,t+\delta t)\approx
   \sum_{n=0}^{4}\frac{(-i\delta t)^n}{n!}
   \left[\hat H_{\mathrm{KS}}\!\left(t+\tfrac{\delta t}{2}\right)\right]^{\!n}
   \psi_{k}(\mathbf r,t).
\tag{7}
\end{equation}
The orbitals are propagated for $N$ time steps up to
$t_{\mathrm{final}} = N\Delta t$, where the time step $\Delta t$ is 
1~attosecond. The small timestep was chosen to ensure numerically stable and accurate time propagation.
The $t_{\mathrm{final}}$ is set to 300 femtosecond for all four
systems and therefore all trajectories are propagated for a total of 300 fs.

A uniform $100\times100\times100$ cubic grid with a spacing of
0.30~Å (a $30.0\times30.0\times30.0$~Å$^{3}$ box) represents the orbitals in
real space.  A 1~as time step and a 0.30~Å grid spacing have been shown to yield
accurate results in previous dissociation dynamics studies.\cite{Taylor2025PRA1,Taylor2025PRA2,Jiang2025JCP,Jiang2026PRA,Viveiros2026PRA}

The Kohn–Sham orbitals are set to zero at the box boundaries.  To prevent
unphysical reflections of electron density when fragments reach the edge, a complex absorbing
potential (CAP) surrounds the box.  Our simulations employ the form proposed by
Manolopoulos\cite{Manolopoulos2002JCP}, applied independently along each Cartesian direction:
\begin{equation}
-iw(x)= -\,\frac{i}{2}
        \Bigl(\frac{2\pi}{\Delta x}\Bigr)^{2}
        f(y),\qquad
y=\frac{x-x_{1}}{\Delta x},\quad y\in[0,1],
\tag{8}
\end{equation}
where $x_{1}$ and $x_{2}=x_1+\Delta x$ are the inner and outer boundaries of the absorbing region along that direction,
$\Delta x=x_{2}-x_{1}$ is the absorbing-layer width, and $f(y)$ is given by
\begin{equation}
f(y)=\frac{4}{c^{2}}
      \left(\frac{1}{(1+y)^{2}}
            +\frac{1}{(1-y)^{2}}-2\right),\qquad c=2.62.
\tag{9}
\end{equation}

Ionic motion is treated classically within the Ehrenfest approximation.  The
force on ion $i$ due to all other ions and the electron density is
\begin{equation}
M_{i}\frac{d^{2}\mathbf R_{i}}{dt^{2}}=
  \sum_{j\neq i}^{N_{\text{ion}}}
      \frac{Z_{i}Z_{j}(\mathbf R_{i}-\mathbf R_{j})}
           {|\mathbf R_{i}-\mathbf R_{j}|^{3}}
  \;-\;
  \int \rho(\mathbf r,t)\,
       \nabla_{\mathbf R_{i}}
       V_{\mathrm{ion}}(\mathbf r;\mathbf R_{i})\,
       d\mathbf r,
\tag{10}
\end{equation}
where $M_{i}$, $Z_{i}$, and $\mathbf R_{i}$ are the mass, pseudocharge
(valence), and position of ion $i$, respectively, and
$N_{\text{ion}}$ is the total number of ions.  The first term is the direct ion--ion Coulomb repulsion (in atomic units), and the second term is the electron-mediated force derived from the pseudopotential.  Equation~(10) is integrated with
the Velocity Verlet algorithm at every time step $\Delta t$.

Because Ehrenfest dynamics uses a mean-field description, it does not capture electronic decoherence or explicit branching between competing quantum pathways; the present results should therefore be interpreted as ensemble-averaged strong-field dynamics.

{\color{black}Nevertheless, the approach has been widely used for strong-field fragmentation and ultrafast proton dynamics because it provides a computationally efficient description of the coupled electron--nuclear motion while retaining the essential feedback between ionization and nuclear rearrangement.}

The initial structures of $(\mathrm{H_2O})_n$ ($n=1$--4) were taken directly from previously reported equilibrium (global-minimum) geometries,\cite{Hoy1979JMS,Jurecka2006PCCP,Maheshwary2001JPCA,Rakshit2019JCP} without further geometry optimization in the present work. The dimer adopts the hydrogen-bonded donor-acceptor minimum, the trimer the cyclic minimum, and the tetramer the cyclic square minimum. Each water cluster was then placed at the center of the simulation box to maximize the available distance to the box boundaries.

\begin{figure}[H]
  \centering
 \includegraphics[width=0.5\textwidth]{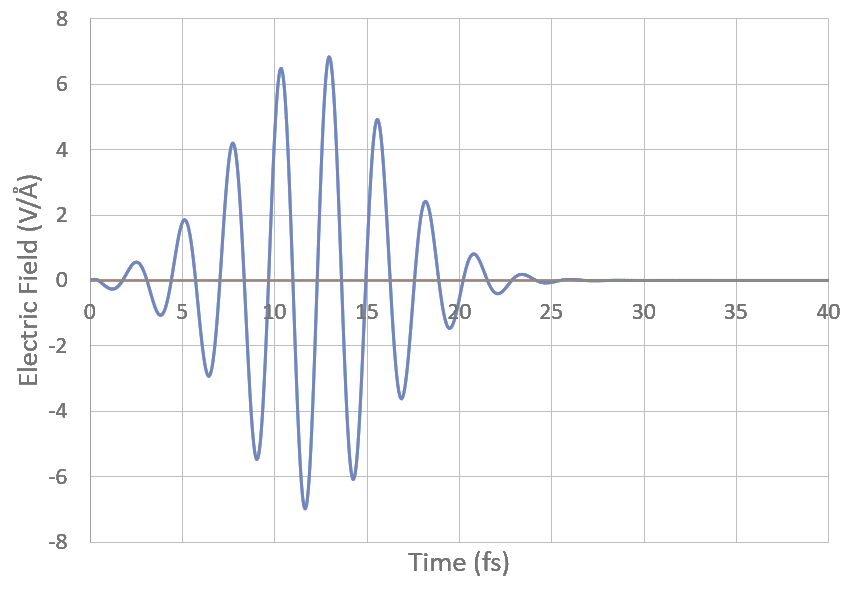}
  \caption{Electric field (unit: V/\AA) versus time (unit: fs) for the laser applied in the simulations.}
  \label{laser}
\end{figure}

We apply a linearly polarized near-infrared few-cycle laser pulse with a central wavelength of
790~nm and a pulse duration of 6~fs (FWHM). The peak field strength is set to
$E_{\max}=7~\mathrm{V/}$\AA. FIG.~\ref{laser} shows the electric field from 0 to 40~fs.
The field is turned on at $t=0$~fs, reaches its maximum amplitude at
$t\approx 12\text{--}13$~fs, and decays to near zero by $t\approx 25$~fs.
These laser parameters were chosen to target an ensemble-averaged net ionization of
approximately two electrons (i.e., an effective charge state of $q\approx 2$) in the water dimer,
which is close to the experimentally relevant charge window in which direct Coulomb-explosion
breakup of the water dimer has been observed.\cite{Zhang2019PRA}
The same laser was then applied to the monomer, dimer, trimer, and tetramer so that
differences in ionization and fragmentation dynamics could be attributed primarily to cluster size,
rather than to changes in the pulse conditions.
{\color{black}Using identical pulse parameters for all cluster sizes ensures that observed differences primarily reflect changes in hydrogen-bond topology rather than changes in the external driving field.}

To mimic an isotropic gas-phase ensemble, the polarization direction $\hat{\mathbf{e}}$ is randomized independently for each trajectory.
Specifically, $\hat{\mathbf{e}}$ is sampled uniformly on the unit sphere by drawing $\phi\in[0,2\pi)$ uniformly and $u=\cos\theta\in[-1,1]$ uniformly, and setting $\hat{\mathbf{e}}=(\sqrt{1-u^{2}}\cos\phi,\sqrt{1-u^{2}}\sin\phi,u)$.
Moreover, initial ionic velocities are sampled randomly from a Boltzmann distribution at
300~K to simulate the random nuclear motions.  This approach allows diverse fragmentation channels and provides a
holistic view of the fragmentation dynamics. {\color{black}The ensemble sizes reflect a compromise between statistical sampling and computational cost: we performed 42 trajectories for the monomer, 104 for the dimer, 93 for the trimer, and 30 for the tetramer. Larger ensembles were used for the dimer and trimer because these systems provide the main quantitative comparison in this work, whereas the tetramer primarily serves to establish the trend at larger cluster size.}

Because the Kohn–Sham density is a continuous function and electron density that flows beyond the CAP boundary is removed from the simulation box, the net ionization computed at the end of the simulation is generally a non-integer number. This is a well-known feature of grid-based TDDFT with absorbing boundaries: the total electron count decreases continuously as ionized density is absorbed, and the final value represents the ensemble-averaged ionization rather than a definite integer charge state. In practice, the non-integer values are interpreted statistically: if a set of otherwise identical trajectories yielded an average ionization of, say, 1.7 electrons, individual members of that set would produce integer charge states distributed around that mean.
{\color{black}Accordingly, the reported ionization values should be interpreted as ensemble-averaged observables rather than integer charge states for any single trajectory.}

In the following section, data including time for H-ejection and H-transfer, end-point connectivity classes, and kinetic energy release (KER) will be given. We present below a definition for each of these quantities.

\paragraph{H-ejection}
An H-ejection event is identified when the distance between a hydrogen atom and its nearest oxygen
first exceeds $3.0$~\AA. This threshold is chosen to be much larger than the equilibrium O--H bond length in
neutral water ($0.957$~\AA). The H-ejection time is defined as the first time at which this $3.0$~\AA\ threshold is reached.

In very rare cases (fewer than $1\%$ of all trajectories), a nearly dissociating O--H unit or
H$_2$O fragment reaches the complex absorbing potential (CAP) before the clean $3.0$~\AA\
criterion is satisfied. In such cases, we still count the hydrogen as ejected if its distance to the
nearest oxygen exceeds $2.5$~\AA\ at the first frame which the atom enters the CAP. 
Also, in those cases the H-ejection time is then defined
as the time when an atom enters the CAP.

\paragraph{H-transfer}

An H-transfer event is identified when a hydrogen atom is separated from its parent oxygen by more
than $2.0$~\AA\ and reaches a different oxygen within $1.5$~\AA, provided that it remains within
$1.75$~\AA\ of the new oxygen for the subsequent 10~fs.
This definition is designed to identify stable H-rearrangement events while excluding transient
proton sharing and transfer-and-return motions.
The H-transfer time is defined as the first time at which the hydrogen reaches within
$1.5$~\AA\ of the new oxygen and subsequently satisfies the 10-fs stability condition.

{\color{black}Because proton motion develops continuously rather than discontinuously, the times reported here should be interpreted as the first times at which the corresponding operational criteria are satisfied, not as the earliest possible initiation times of the underlying dynamics.}

\paragraph{Kinetic energy release (KER).}
We evaluate the KER only for the direct two-body dissociation subset of the water dimer,
i.e., trajectories that dissociate into two H$_2$O fragments.
This is the only channel in the present dataset for which a clean
experimental value is
available, and it is also the most reliable subset for KER analysis.
For other fragmentation channels, especially in the trimer and tetramer, fragments frequently
approach the complex absorbing potential (CAP) and may lose electronic density before a clean
late-time energy analysis can be carried out, making the corresponding KER estimates less robust.

For each accepted dimer trajectory, we first compute the translational kinetic energy of the two
$\mathrm{H_2O}$ fragments from their center-of-mass velocities at the final time $t_f$,
\begin{equation}
\mathrm{KER}_{\mathrm{COM}}(t_f)
=
\frac{1}{2} M_{\mathrm{H_2O}} \left| \mathbf{V}^{\mathrm{CM}}_1(t_f) \right|^2
+
\frac{1}{2} M_{\mathrm{H_2O}} \left| \mathbf{V}^{\mathrm{CM}}_2(t_f) \right|^2,
\tag{11}
\end{equation}
where $M_{\mathrm{H_2O}}$ is the mass of one water molecule and $\mathbf{V}^{\mathrm{CM}}_i(t_f)$ is the center-of-mass velocity
of fragment $i$ at $t_f$.

Because the two charged fragments are still at a finite separation at $t_f$, part of the Coulomb
potential energy has not yet been converted into translational kinetic energy.
We therefore estimate the asymptotic KER as
\begin{equation}
\mathrm{KER}_{\infty}
\approx
\mathrm{KER}_{\mathrm{COM}}(t_f)
+
\frac{q_1 q_2 e^2}{4 \pi \varepsilon_0 R_{12}(t_f)},
\tag{12}
\end{equation}
where $q_1$ and $q_2$ are the net charges of the two fragments (in units of the elementary charge $e$), $R_{12}(t_f)$ is the distance between the two fragment centers of mass at $t_f$, and the Coulomb prefactor $e^2/(4\pi\varepsilon_0)$ converts the 
result to eV when $R_{12}$ is in \AA\  (using $e^2/(4\pi\varepsilon_0)\approx14.3996$~eV\,\AA).
Thus, the reported dimer KER corresponds to the final translational center-of-mass kinetic energy
plus the residual Coulomb potential energy remaining at the end of the simulation.
{\color{black}The fragment charges used in Eq.~(12) are obtained from the final integrated electronic density associated with each oxygen-containing fragment.}

In the next section, we present the size-dependent ionization and dissociation dynamics of
(H$_2$O)$_n$ ($n=1$--4) under the applied few-cycle laser pulse.
Specifically, we (i) summarize the ionization statistics across cluster sizes to establish the baseline electronic response,
(ii) show representative trajectory snapshots for the dimer, trimer, and tetramer to illustrate the real-space structural evolution,
(iii) analyze the probability and timing of H-ejection,
(iv) analyze the probability and timing of H-transfer,
(v) characterize the endpoint oxygen response through the dimer O--O separation distribution and the coarse endpoint connectivity classes of the trimer and tetramer, and
(vi) compare our simulation results with the available experimental studies.

\section{\label{sec:level3}Results and Discussion}

\subsection{\label{sec:level4}Ionization Statistics}

\begin{figure}[H]
  \centering
  \includegraphics[width=0.5\textwidth]{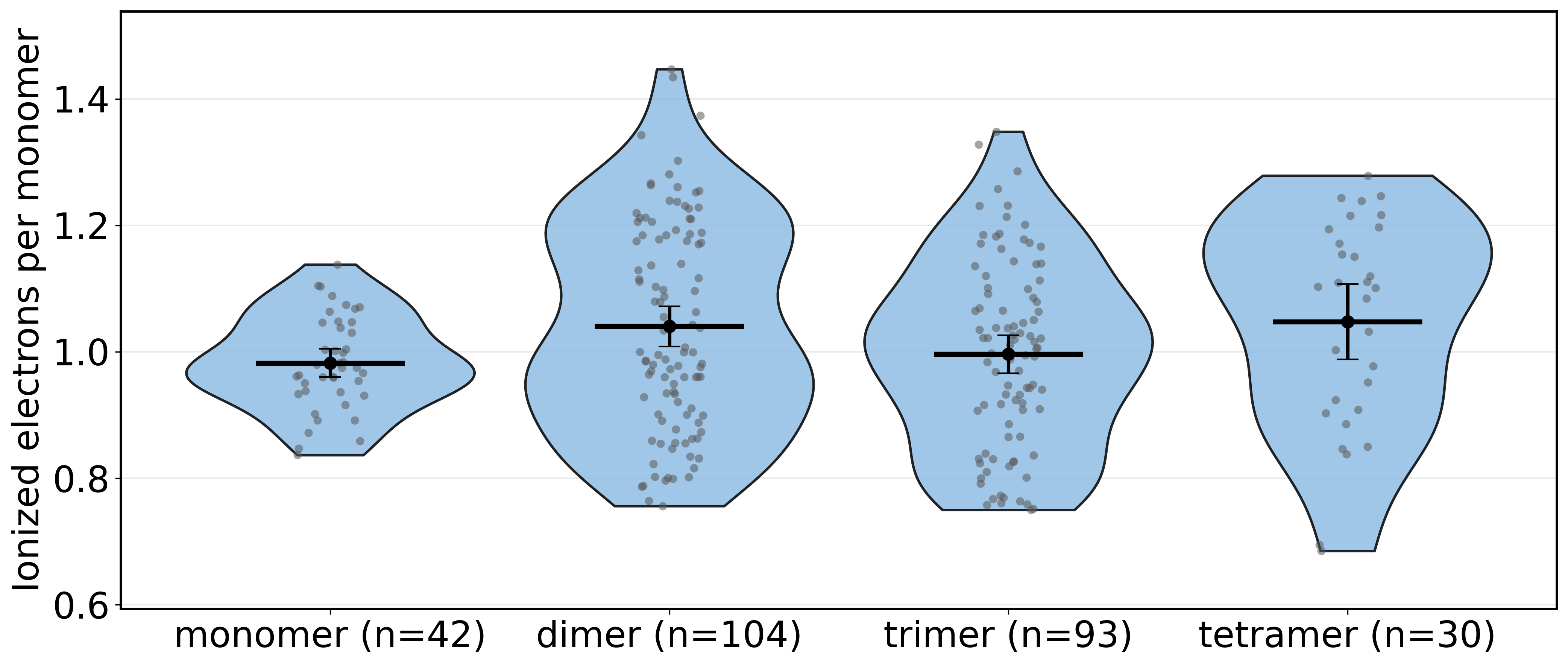}
  \caption{
Net ionization per monomer at $t = 25$~fs (after the laser field has
decayed to near zero) for (\ce{H2O})$_n$ ($n = 1$--$4$), evaluated
over 42, 104, 93, and 30 trajectories, respectively.
Each point represents one trajectory; values are jittered
horizontally for visibility.
The horizontal line marks the mean and error bars
denote 95\% confidence intervals.
Mean values are 0.98, 1.04, 1.00, and 1.05
electrons per monomer for $n = 1$--$4$, respectively. The pairwise differences in the means are
small, with the overall spread across all four systems less than 8\% of the mean, indicating that net ionization per monomer is
largely insensitive to cluster size under the present pulse conditions.
Broader, more asymmetric distributions for
larger clusters reflect the increasing heterogeneity of accessible
ionization pathways.  
}
  \label{ionization}
\end{figure}

FIG.~\ref{ionization} shows the distribution of the ionized electrons per monomer for
(H$_2$O)$_n$ ($n = 1$--4). The ionization is evaluated at $t = 25$~fs, when the laser field has
essentially vanished. As shown later in FIG.~\ref{e_time} for the dimer ensemble, more than 99\%
of the final net ionization is already complete by $\sim$17.5~fs, so the value at 25~fs represents a
fully converged plateau rather than a snapshot during active ionization; the same convergence is
expected for all four cluster sizes under identical pulse conditions. The per-monomer quantity is
used to place all four systems on a common footing independent of cluster size.

\textit{Overall trend.}
The four systems do not exhibit a strong monotonic size dependence in the mean ionization per
monomer. The mean values are 0.9823, 1.0402, 0.9963, and 1.0478~e per monomer for the
monomer, dimer, trimer, and tetramer, respectively, all clustered within a range of only
0.07~e --- a variation of less than 8\% relative to the mean. These per-monomer values imply
total ensemble-averaged charge states of approximately $q \approx 1.0$, 2.1, 3.0, and 4.2
electrons for the monomer through tetramer, respectively, consistent with the laser parameters
having been chosen to target $q \approx 2$ in the dimer (Sec.~\ref{sec:level2}). The
corresponding distributions show substantial overlap across all four systems, and the pairwise differences in the means are small relative to the spread of individual-trajectory values,
as is evident from FIG.~\ref{ionization}. Under the present pulse
conditions, increasing cluster size does not by itself lead to a dramatic change in the net
ionization per monomer.

\textit{Distribution shapes.}
Beyond the means, the violin plots reveal systematic differences in the shape and spread of
the ionization distributions across cluster sizes. The monomer distribution is the narrowest
and most symmetric of the four, reflecting the absence of any hydrogen-bond-network effects
and the relatively uniform strong-field response of a single isolated molecule sampled across
42 random orientations. The dimer distribution is broader and develops a slight upward tail, consistent with a wider spread of per-monomer ionization outcomes once the donor-acceptor asymmetry of the single hydrogen bond is included.

The trimer distribution is similarly broad, spanning roughly 0.75--1.35~e per monomer with a somewhat asymmetric spread. The breadth of the trimer may reflect the enhanced orientation sensitivity of the cyclic-ring geometry: the
cyclic trimer presents three distinct O--H bond orientations with respect to the laser polarization
axis, and the ring can be tilted at a wide range of angles relative to the field, producing a
correspondingly wider spread in the number of bonds that are efficiently ionized in any given
trajectory. The tetramer distribution is also broad, with its spread reflecting both genuine physical heterogeneity and the more limited statistical sampling (30 trajectories), which reduces suppression of extreme outcomes relative to the
larger dimer and trimer ensembles (104 and 93 trajectories, respectively).

\textit{Odd--even pattern.}
Comparing the four means, the monomer and trimer are very close to each other (0.9823 and 0.9963~e per monomer), as are the dimer and tetramer (1.0402 and 1.0478~e per monomer), with both even-membered clusters ionizing slightly more per monomer than their odd-membered neighbors. This suggests a possible odd--even trend, but the present dataset is not sufficient to establish it robustly: the confidence intervals overlap substantially, the sample sizes are uneven, and a definitive test would require a larger ensemble with orientation-resolved analysis.

\textit{Baseline characterization.}
Taken together, these results establish the net ionization per monomer as a weakly
size-dependent baseline for the present study. The modest spread in means (less than 8\%
across all four systems) and the small pairwise differences relative to the within-group spread
confirm that the dramatic size-dependent effects reported in the following sections --- a
more-than-fourfold increase in H-ejection activity from the dimer to the trimer, and an
approximately eightyfold increase in stable H-transfer over the same step --- cannot be
attributed to differences in total charge deposition. Instead, they point to an important role
of the hydrogen-bond-network topology and the multi-center nature of the Coulomb interactions
in a multiply ionized cluster, which reshape the protonic and structural response far more
strongly than the net electron removal itself.

\subsection{\label{sec:level3B}Representative Dissociation Dynamics of the Dimer, Trimer,
and Tetramer}

\begin{figure}[H]
\centering
\includegraphics[width=0.5\textwidth]{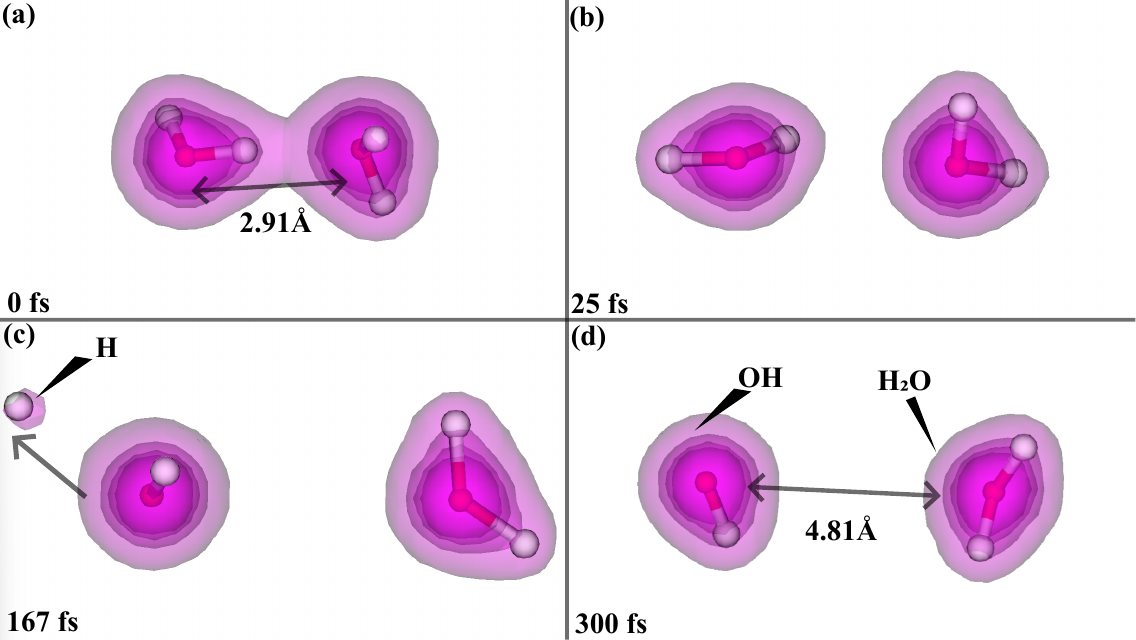}
\caption{Representative snapshots along a three-body dissociation
trajectory of the H$_2$O dimer,
ultimately yielding H + OH + H$_2$O. The initial O--O distance is
2.91~\AA\ [panel (a)] and
increases to 4.81~\AA\ at 300~fs [panel (d)]. An H-ejection event
occurs at 167~fs from one of
the two water monomers. Four electron-density isosurfaces are
displayed in each panel. The
isosurface values are 0.10, 0.567, 1.033, and 1.50.}
\label{dimer}
\end{figure}

\begin{figure}[H]
\centering
\includegraphics[width=0.5\textwidth]{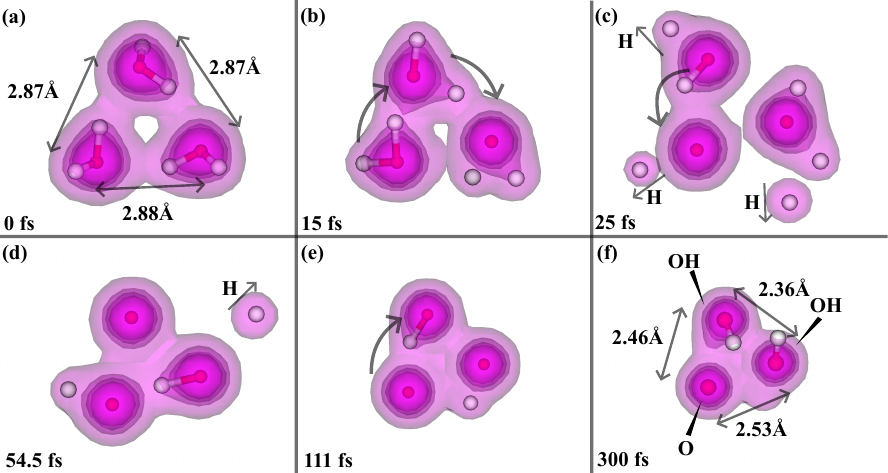}
\caption{
Representative snapshots along a partially dissociating
trajectory of the H$_2$O trimer.
The oxygen atoms are labeled O1 (top), O2 (bottom left), and
O3 (bottom right).
The initial O--O distances are 2.87, 2.87, and 2.88~\AA\ for
O1--O2, O1--O3, and O2--O3,
respectively, and the corresponding final separations are
2.46, 2.36, and 2.53~\AA.
The trajectory contains four H-ejection events and two
H-transfer events, and ultimately leaves a OH--OH--O complex
after partial dissociation.
Four electron-density isosurfaces are displayed in each
panel. The isosurface values are 0.10, 0.567,
1.033, and 1.50.
}
\label{trimer}
\end{figure}

\begin{figure}[H]
\centering
\includegraphics[width=0.5\textwidth]{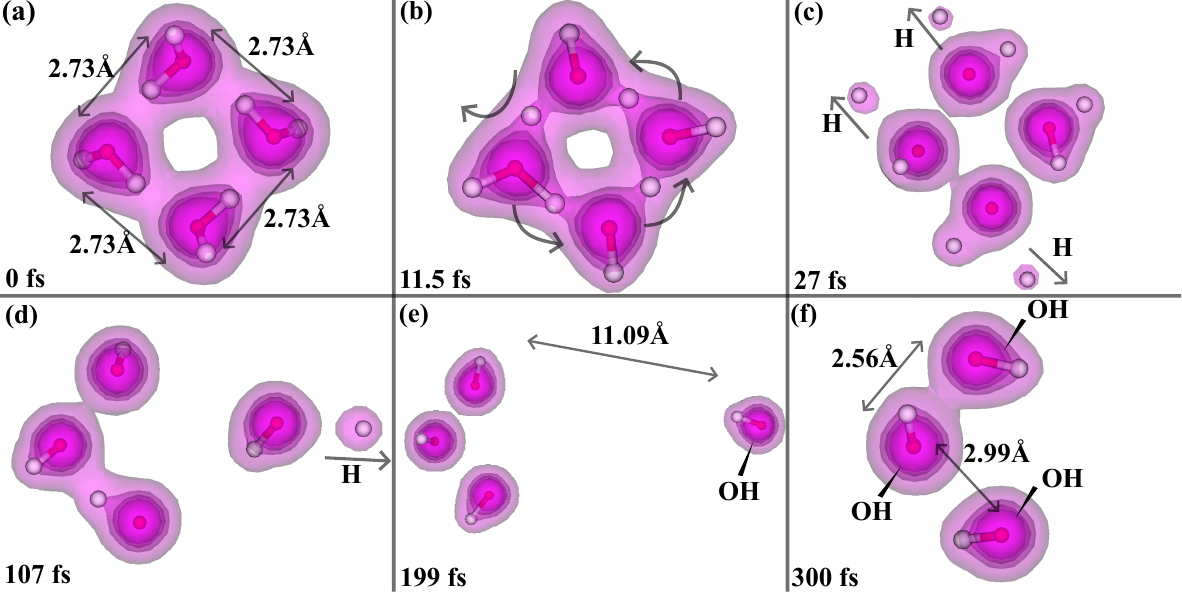}
\caption{
Representative snapshots along a partially
dissociating trajectory of the H$_2$O tetramer.
The oxygen atoms are labeled O1 (top), O2 (left), O3
(right), and O4 (bottom).
The initial O--O distances are 2.73~\AA\ for all four
neighboring oxygen pairs.
At later times, O3 reaches the CAP at 199~fs [panel
(e)], so only the remaining well-defined
neighboring separations are reported at the final
time: 2.56 and 2.99~\AA\ for O1--O2 and O2--O4,
respectively.
The trajectory contains four H-ejection events and
three H-transfer events, and ultimately leaves a
loosely bound three-oxygen complex together with a
departing OH fragment.
Four electron-density isosurfaces are displayed in
each panel. The isosurface values are 0.10, 0.567,
1.033, and 1.50.
}
\label{tetramer}
\end{figure}

FIG.~\ref{dimer}--FIG.~\ref{tetramer} show
representative dissociation snapshots of the dimer,
trimer, and tetramer trajectories. For the water
monomer, only one out of 42 trajectories
dissociates into H + OH within 300~fs, so no
representative monomer snapshot is shown.

\textit{Dimer.}
The representative dimer trajectory illustrates
several features that are characteristic of the
dimer ensemble as a whole.
During the laser pulse [FIG.~\ref{dimer}(a)--(b)],
the electron density isosurfaces deform as ionization
proceeds, and the nuclear framework begins to respond: the O--O
distance shows an early increase in this window,
indicating that the nuclear response is already underway on the sub-25-fs timescale.
The protonic activity in this trajectory is predominantly
post-pulse: a single H-ejection event occurs at
167~fs [FIG.~\ref{dimer}(c)], more than 140~fs after
the laser field has essentially vanished.
This delayed response is physically consistent with
the picture that, in the dimer, the
multi-ionization charge buildup drives a slow
Coulomb-assisted dissociation of the O--H bond
rather than a rapid field-driven expulsion during the pulse.
Accompanying the H-ejection, the O--O distance
expands substantially from 2.91~\AA\ at
$t=0$ to 4.81~\AA\ at 300~fs [FIG.~\ref{dimer}(d)],
an increase of nearly 1.9~\AA, reflecting
the Coulomb repulsion between the two positively
charged oxygen-containing fragments once the
bridging hydrogen is lost.
The trajectory ultimately yields a three-body H + OH
+ H$_2$O endpoint. This channel
represents a minority but still significant fraction
of the dimer ensemble, since approximately
30\% of dimer trajectories exhibit at least one
H-ejection event, and it provides a concrete
real-space illustration of the strong coupling
between H-ejection and O--O expansion that is
examined statistically in Sec.~\ref{sec:level5}.

\textit{Trimer.}
The representative trimer trajectory exhibits
qualitatively richer and faster protonic dynamics
than the dimer example, with the most striking
difference being that the bulk of the protonic
activity is already complete by the end of the laser pulse.
At 15~fs [FIG.~\ref{trimer}(b)], midway through the
pulse, the electron density has already
substantially reorganized: the isosurfaces show
visible elongation of several O--H bonds, signaling
that ionization-driven bond weakening is
well underway while the field is still active.
By 25~fs [FIG.~\ref{trimer}(c)], when the laser
field has essentially vanished, three H-ejection
events have already occurred --- a striking contrast
with the dimer, where no H-ejection is
observed until 167~fs. A fourth H-ejection follows
at 54.5~fs [FIG.~\ref{trimer}(d)], close on
the heels of the pulse. The trajectory also contains
two stable H-transfer events: between
panels (b) and (c), one hydrogen migrates from O1 to
O3, while a second hydrogen undergoes
a back-and-forth excursion between O1 and O2 that
does not satisfy the stability criterion and
therefore does not count as a transfer. This
transient proton-sharing motion illustrates the
competition between ejection, stable transfer, and
transfer-and-return that is inherent to the
multi-ionized hydrogen-bond network. A second stable
H-transfer occurs at 111~fs
[FIG.~\ref{trimer}(e)], after the pulse has ended.

A counterintuitive but noteworthy feature of this
trajectory is that, despite four H-ejection events
and two H-transfer events, the oxygen framework does
not expand but instead \textit{compresses}.
The three O--O distances decrease from their initial
values of 2.87, 2.87, and 2.88~\AA\ to final
values of 2.46, 2.36, and 2.53~\AA, respectively ---
a contraction of roughly 0.35--0.51~\AA\ on
each pair. 
One possible explanation is that
proton rearrangement partially redistributes positive charge
across the remaining oxygen-containing fragments, altering
the balance between attractive and repulsive interactions for
specific oxygen pairs.
The trajectory ultimately retains a loosely bound
OH--OH--O complex, and this
compressed-endpoint behavior is not exceptional: it
is observed in a substantial fraction of the
trimer ensemble, as quantified in Sec.~\ref{sec:level6E}.

\textit{Tetramer.}
The representative tetramer trajectory shows the
most intense and earliest protonic response of
the three systems, consistent with its shorter
initial O--O separations (2.73~\AA\ for all
neighboring pairs, compared with 2.87--2.88~\AA\ in
the trimer and 2.91~\AA\ in the dimer).
The shorter inter-oxygen distances lower the barrier
to proton migration and place donor
hydrogens geometrically closer to neighboring
acceptor oxygens at the onset of the pulse,
facilitating faster and more widespread H-transfer initiation.

Already at 11.5~fs [FIG.~\ref{tetramer}(b)], during
the pulse, three simultaneous H-transfer
motions are initiated: donor hydrogens on the O1--,
O2--, and O4-containing units are each
driven toward their respective acceptor oxygens.
Notably, the hydrogen between O1 and O2 is
marked with an outward arrow in the figure because,
although initially displaced toward O2, it is
subsequently diverted and ultimately ejected rather
than completing a stable transfer. This
diversion illustrates a key dynamical competition in
the tetramer: field-driven proton motion
can be redirected mid-trajectory by the rapidly
evolving charge distribution, so that early
transfer-like displacement does not guarantee a
stable rearrangement outcome.

The ejection response in this trajectory is both
early and extensive. By 27~fs
[FIG.~\ref{tetramer}(c)], only 2~fs after the laser
has essentially vanished, three hydrogens
have already been ejected from the O1--, O2--, and
O4-containing units. This is in stark
contrast with the dimer, where the single H-ejection
does not occur until 167~fs. A fourth
H-ejection from the O3-containing unit follows at
107~fs [FIG.~\ref{tetramer}(d)], by which
time O3 is already well separated from the remaining
three-oxygen core.

The subsequent oxygen-framework evolution reveals an
asymmetric partial-breakup pathway.
At 199~fs [FIG.~\ref{tetramer}(e)], the OH fragment
containing O3 reaches the CAP boundary,
with an O1--O3 separation of 11.09~\AA\ --- a
separation more than four times the initial
O--O distance, confirming that O3 has undergone a
full Coulomb-driven departure. Meanwhile,
the remaining three-oxygen core stays comparatively
compact. By 300~fs
[FIG.~\ref{tetramer}(f)], the O1--O2 and O2--O4
separations are only 2.56 and 2.99~\AA,
respectively, indicating that the three residual
oxygen-containing units remain loosely bound as
a (OH)$_3$ complex while a single OH fragment
departs. This trimer + monomer endpoint is
one of the most common tetramer outcomes in the
present dataset, as shown in
Sec.~\ref{sec:level6E}.

\textit{Cross-system comparison.}
Taken together, the three representative
trajectories reveal a clear and systematic evolution of
the strong-field response with increasing cluster
size. The most prominent trend is the shift in
the \textit{timing} of protonic activity: in the
dimer, both H-ejection and O--O expansion are
entirely post-pulse phenomena, whereas in the trimer
and tetramer the dominant protonic
response --- including multiple H-ejections and the
initiation of H-transfer --- is already
underway or complete within the laser pulse window.
This timing shift is consistent with a network-driven mechanism in which the extended hydrogen-bond topology channels ionization-induced forces into rapid, cooperative proton motion during the pulse, rather than the slower, Coulomb-assisted post-pulse response seen in the dimer.
A second notable trend is the complex relationship between
protonic activity and oxygen-framework
response: the dimer shows large O--O expansion
accompanying its single delayed H-ejection,
whereas the trimer in the representative trajectory
undergoes extensive protonic rearrangement
while its oxygen framework actually compresses. The
tetramer, in turn, exhibits a hybrid
behavior --- the oxygen framework partially breaks
up through the departure of a single OH
fragment, while the remaining three-oxygen core
stays bound. These snapshots are qualitative
illustrations; the following sections quantify these
trends systematically across the full trajectory
ensembles.

\subsection{\label{sec:level5}H-ejection}

\begin{figure}[H]
\centering
\includegraphics[width=0.5\textwidth]{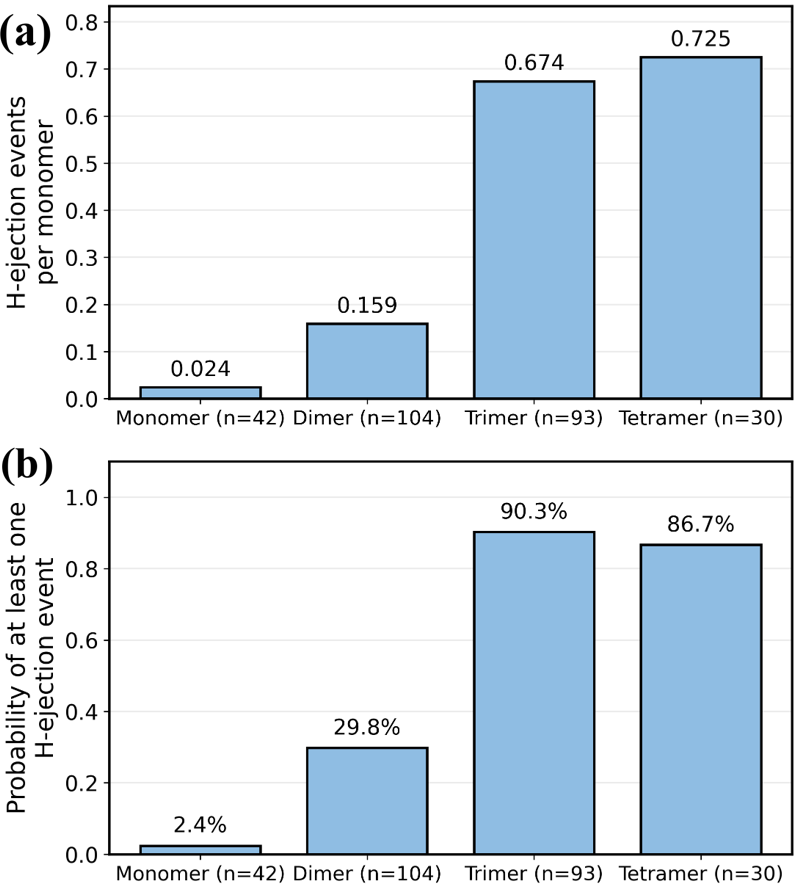}
\caption{
Bar plots of (a) the number of H-ejection
events per monomer and (b) the probability of
at least
one H-ejection event across (H$_2$O)$_n$ ($n=1$--4).
The H-ejection events per monomer are
calculated as the total number of H-ejection
events across
all trajectories divided by the total number
of monomers in those trajectories.
The probability of at least one H-ejection
event is calculated as the fraction of
trajectories in which
at least one H-ejection event occurs.
}
\label{eject_prob}
\end{figure}

FIG.~\ref{eject_prob} summarizes the H-ejection statistics across (H$_2$O)$_n$ ($n=1$--4) in
terms of both the number of H-ejection events per monomer [panel (a)] and the probability of
observing at least one H-ejection event in a trajectory [panel (b)].

\textit{Monomer.}
The monomer shows essentially no H-ejection under the present conditions: only one out of 42
trajectories exhibits an H-ejection event, occurring at roughly 230~fs, giving a very low event
rate per monomer (0.024) and a probability of at least one H-ejection event of only 2.4\%.
The extremely late timing of this single event --- more than 200~fs after the laser field has
vanished --- indicates that even the rare monomer H-ejection is not field-driven but instead
reflects a slow, Coulomb-assisted dissociation of a highly ionized O--H bond on the ground-state
repulsive surface of the multiply charged monomer ion. The near-complete suppression of
H-ejection in the monomer under the present conditions establishes an important baseline:
the strong H-ejection activity observed in the larger clusters is a network effect rather than
a property of the individual water molecule.

\textit{Dimer.}
The dimer already shows a clear increase relative to the monomer, with 0.159 H-ejection
events per monomer and a 29.8\% probability of at least one H-ejection event. The more-than-sixfold increase in the event rate relative to the monomer (0.024 $\to$ 0.159) confirms that
the hydrogen-bond partner substantially enhances H-ejection propensity, even though the dimer
ionization per monomer is only marginally higher than that of the monomer (Sec.~\ref{sec:level4}).
This decoupling between ionization and H-ejection already signals that network topology,
rather than total charge deposition alone, governs the protonic response.

\textit{Trimer and tetramer.}
A much stronger size effect emerges beyond the dimer. Both the trimer and tetramer display
very high H-ejection activity, with 0.674 and 0.725 events per monomer, respectively, and
probabilities of at least one H-ejection event of 90.3\% and 86.7\%. The jump from the dimer
to the trimer alone represents a more than fourfold increase in the event rate per monomer
(0.159 $\to$ 0.674) and a tripling of the per-trajectory probability (29.8\% $\to$ 90.3\%),
underscoring that the dimer--trimer transition is the dominant step in the size dependence of
H-ejection. This stands in sharp contrast to the net ionization per monomer, which changes
by less than 5\% across the same step (Sec.~\ref{sec:level4}), confirming that the H-ejection
size effect is governed primarily by the hydrogen-bond-network topology rather than by
increased charge deposition.

A further distinction between the trimer and tetramer is
worth noting, though it should be interpreted cautiously given the limited tetramer sample size (30 trajectories): the tetramer has a
slightly higher event rate per monomer (0.725 vs.\ 0.674) but a slightly lower per-trajectory
probability (86.7\% vs.\ 90.3\%). This combination suggests that when H-ejection occurs in
the tetramer it tends to involve more ejection events per trajectory on average, but a small
fraction of tetramer trajectories produce no H-ejection at all.

\begin{figure}[H]
\centering
\includegraphics[width=0.5\textwidth]{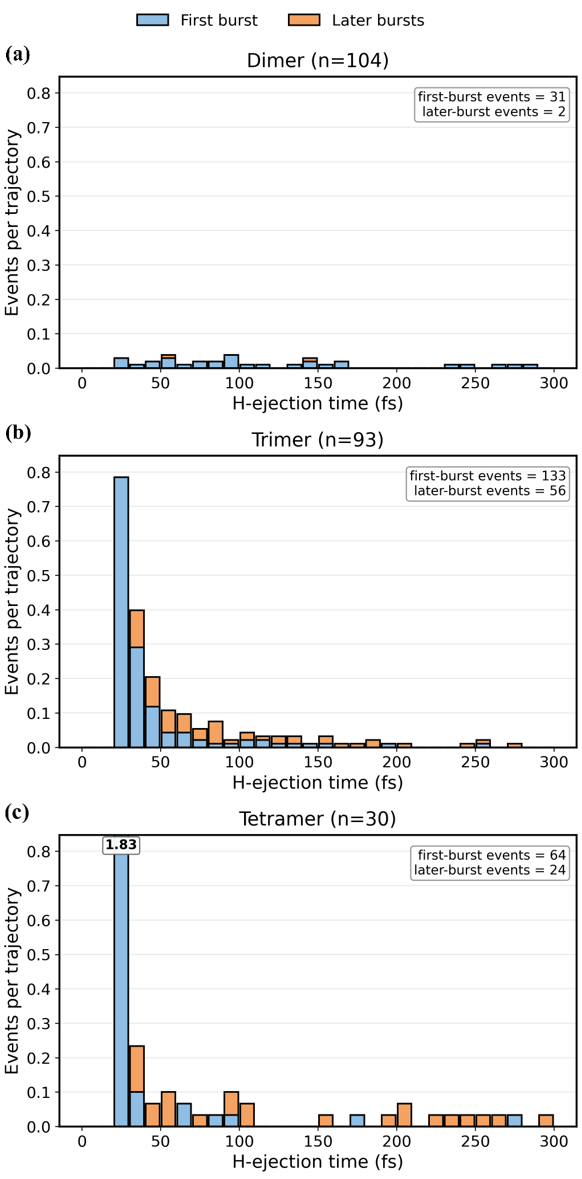}
\caption{
Histograms showing the time distribution of H-ejection events for the dimer,
trimer, and tetramer [panels (a)--(c), respectively]. Each bar corresponds to a 10-fs time
interval, and the y-axis gives the number of events per trajectory, calculated as the total
number of H-ejection events occurring in that interval divided by the total number of
trajectories for the corresponding system. The first burst is defined as the first H-ejection
event in a trajectory together with all subsequent H-ejection events occurring within 10~fs
of that first event. All remaining H-ejection events are classified as later-burst events.
First-burst events are shown in blue, and later-burst events are shown in orange.
}
\label{eject_time}
\end{figure}

\textit{Timing distributions.}
FIG.~\ref{eject_time}(a)--\ref{eject_time}(c) show the time distributions of H-ejection events
for the dimer, trimer, and tetramer, plotted as events per trajectory in 10-fs bins. Because the
laser pulse frequently induces multiple H-ejection events within a short time window, the first
burst is defined as the first H-ejection event in a trajectory together with all subsequent
H-ejection events occurring within 10~fs of that first event; all remaining H-ejection events
are classified as later-burst events. This representation separates the earliest pulse-driven
H-ejection activity from weaker delayed secondary emission.

For the dimer [FIG.~\ref{eject_time}(a)], later-burst events are rare: only 2 later-burst
events are recorded across all 104 dimer trajectories, compared with 31 first-burst events.
This means that dimer H-ejection is almost always a single, isolated event rather than a
cascade. The first-burst events are distributed broadly across the simulation window, with
activity spread relatively evenly across the 20--100~fs range (with the 90--100~fs bin representing the single largest contribution), and a non-negligible tail extending to 130--170~fs and a smaller
residual contribution in the 230--290~fs range. Notably, the
onset of H-ejection already at 20--30~fs --- during the trailing edge of the laser pulse
($\sim$25~fs) --- indicates that a fraction of dimer H-ejection is field-assisted rather than
purely post-pulse, nuancing the picture established in Sec.~\ref{sec:level3B}. Nevertheless,
the dominant character of the dimer response is delayed: the distribution is spread over
more than 200~fs with all bins lying below 0.05 events per trajectory per 10-fs bin,
reflecting a slow, thermally and Coulomb-driven process rather than a sharp, field-driven one.

For the trimer [FIG.~\ref{eject_time}(b)], the H-ejection signal is strongly concentrated in
the early-time window. The 20--30~fs bin is the dominant feature, reaching approximately
0.8 events per trajectory and containing only first-burst events, since H-ejection does not
begin until around 20~fs in the trimer dataset. The first-burst-to-later-burst ratio across all
trimer trajectories is approximately 133:56 ($\sim$70:30), indicating that the early cooperative
pulse-driven response is the primary channel, but secondary ejection from already-destabilized
O--H bonds is also significant. The overall H-ejection intensity decreases substantially beyond
50~fs, and beyond 100~fs all bins drop below 0.05 events per trajectory per 10-fs bin ---
comparable to the low-level dimer background at those times. The late-time tail is dominated
entirely by later-burst events, consistent with a picture in which the pulse-driven first ejection
destabilizes the remaining O--H bonds, leading to secondary emission on a slower timescale.

For the tetramer [FIG.~\ref{eject_time}(c)], the early-time peak is even more dramatic. The
20--30~fs bin reaches 1.83 events per trajectory, meaning that on average nearly two H-ejection
events occur per trajectory within this single 10-fs interval alone --- more than twice the
corresponding trimer peak value of $\sim$0.8. The first-burst-to-later-burst ratio across all
tetramer trajectories is approximately 64:24 ($\sim$73:27), similar to the trimer ratio,
but the absolute magnitude of the early-time response is far larger. As a consequence of
this dominant early peak, most trajectories that exhibit any H-ejection at all have already
done so by 30~fs, and the signal from 30 to 300~fs consists almost entirely of later-burst
events. No H-ejection events appear in the 110--150~fs range in the present dataset; although
this gap likely reflects in part the more limited tetramer sample size (30 trajectories), it
may also indicate a genuine lull between the initial fast-ejection cascade and residual
later activity. Beyond 150~fs, a weak late-time tail persists through the 150--300~fs
window. Its level is slightly higher than those of the dimer and trimer, but this difference
should be interpreted cautiously given the smaller sample.

Taken together, the H-ejection statistics reveal two distinct aspects of the size dependence.
In the total event rate and per-trajectory probability, the dominant transition occurs between
the dimer and the trimer, with relatively little further change from the trimer to the tetramer.
In the timing distribution, by contrast, the tetramer is clearly more extreme than the trimer:
its H-ejection is more strongly concentrated in the earliest 10-fs window, with a larger
first-burst peak. Both aspects are consistent with a mechanism in which the extended hydrogen-bond
network of the larger clusters channels ionization-induced charge redistribution into rapid,
cooperative O--H bond rupture during the pulse, driven by the combined effects of the
reduced initial O--O distances and the multi-center nature of the Coulomb repulsion in a
multiply ionized network. By contrast, the dimer response is dominated by isolated, delayed
single-ejection events, consistent with a slower Coulomb-driven dissociation on the
repulsive surface of the charge-separated dimer ion.
We next turn to H-transfer, which provides a more discriminating probe of how proton
rearrangement evolves from the dimer to the larger clusters.

\subsection{\label{sec:level6D}H-transfer}

\begin{figure}[H]
  \centering
  \includegraphics[width=0.5\textwidth]{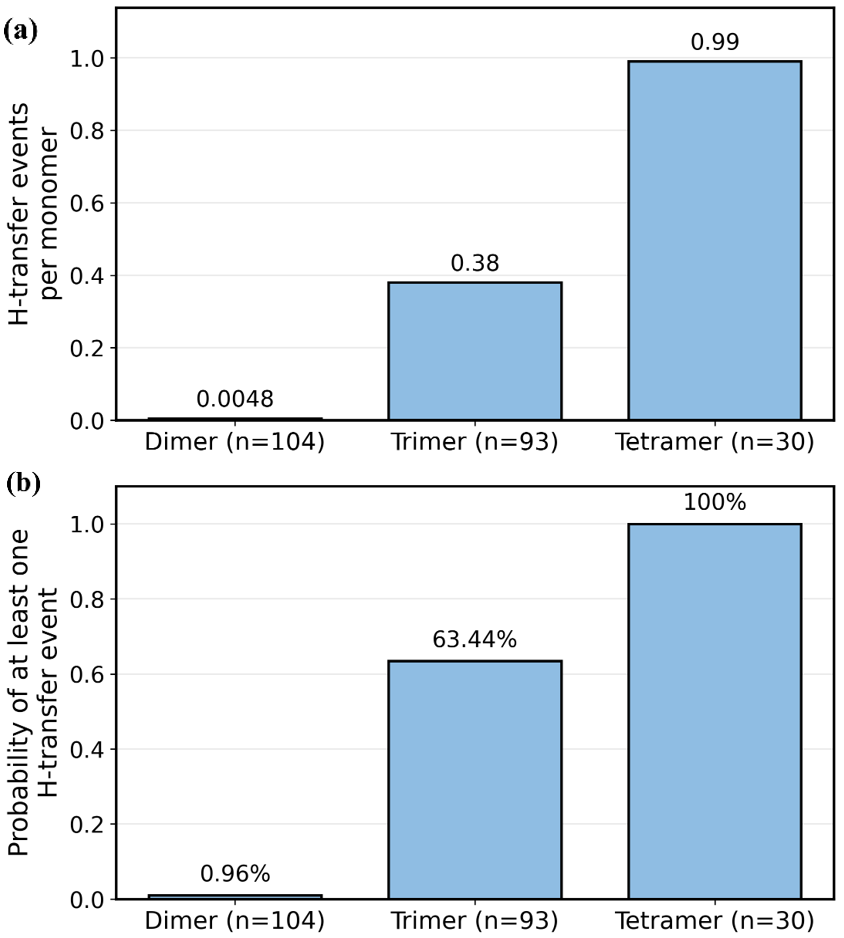}
  \caption{
Bar plots of (a) the number of H-transfer events per monomer and (b) the probability of at least
one H-transfer event across (H$_2$O)$_n$ ($n=2$--4).
The H-transfer events per monomer are calculated as the total number of H-transfer events across
all trajectories divided by the total number of monomers in those trajectories.
The probability of at least one H-transfer event is calculated as the fraction of trajectories in which
at least one H-transfer event occurs.
}
  \label{transfer_prob}
\end{figure}

FIG.~\ref{transfer_prob} summarizes the H-transfer statistics for the dimer, trimer, and tetramer
in terms of both the number of H-transfer events per monomer [panel (a)] and the probability of
observing at least one H-transfer event in a trajectory [panel (b)].

\textit{Dimer.}
Under the present strict H-transfer definition, the dimer shows essentially no stable H-transfer:
only 0.0048 H-transfer events per monomer and a probability of at least one H-transfer event of
only 0.96\%. Across all 104 dimer trajectories, a single H-transfer event is detected, occurring
at 134~fs. This event happens more than 100~fs after the laser field has essentially vanished,
a timing analogous to the rare monomer H-ejection event at $\sim$230~fs, and it therefore
reflects post-pulse Coulomb-driven dynamics on the repulsive surface of the multiply charged
dimer rather than field-driven proton migration. The near-total suppression of stable H-transfer
in the dimer is physically significant: it establishes that, under the present few-cycle pulse
conditions, the effective ionization window is too brief for stable proton rearrangement to complete
before the charge state reaches its final value, consistent
with the ionization dynamics discussed in Sec.~\ref{sec:KER}. It also provides an important
contrast with the longer-pulse experimental conditions of Zhang \textit{et al.}\cite{Zhang2019PRA},
in which a proton-transfer time constant of $31 \pm 5$~fs was inferred --- a timescale fully
comparable to their 38-fs pulse duration but much longer than the effective ionization window
of the present few-cycle pulse.

\textit{Trimer.}
A striking size effect emerges at the trimer. The H-transfer event rate rises to 0.38 per monomer
and the per-trajectory probability reaches 63.44\%, corresponding to an approximately eightyfold
increase in the event rate (0.0048 $\to$ 0.38) and a sixtyfold increase in the probability
(0.96\% $\to$ 63.44\%) relative to the dimer. For context, the corresponding dimer-to-trimer
step in H-ejection was only a fourfold increase in event rate and a threefold increase in
probability (Sec.~\ref{sec:level5}), making H-transfer the far more sensitive indicator of the
network-driven transition at the dimer--trimer boundary. In the trimer's
cyclic geometry, each oxygen simultaneously acts as acceptor to one neighbor and
donor to another, providing multiple alternative migration pathways and a stabilizing network
environment that is absent in the dimer's single donor--acceptor geometry.

\textit{Tetramer.}
The tetramer shows an even more dramatic enhancement. The H-transfer event rate rises to
0.99 per monomer, and every tetramer trajectory in the present dataset exhibits at least one
H-transfer event (100\% probability). The rate increase from the trimer to the tetramer
(0.38 $\to$ 0.99) is approximately 2.6-fold, smaller in ratio than the dimer-to-trimer jump
but substantial in absolute terms, and the saturation of the per-trajectory probability at 100\%
indicates that H-transfer has become a universal feature of the tetramer response under the
present conditions. This saturation is physically consistent with the cyclic-square topology of
the tetramer: all four donor hydrogens face a neighboring acceptor oxygen at the initial O--O
separation of 2.73~\AA, the shortest of any cluster studied here, so that multi-ionization
charge redistribution activates proton migration across the entire ring essentially without
exception. The event rate of $\sim$1 per monomer further suggests that, on average, every
monomer unit in the tetramer participates in at least one H-transfer event over the course
of the trajectory.

\textit{Timing distributions.}

\begin{figure}[H]
  \centering
  \includegraphics[width=0.5\textwidth]{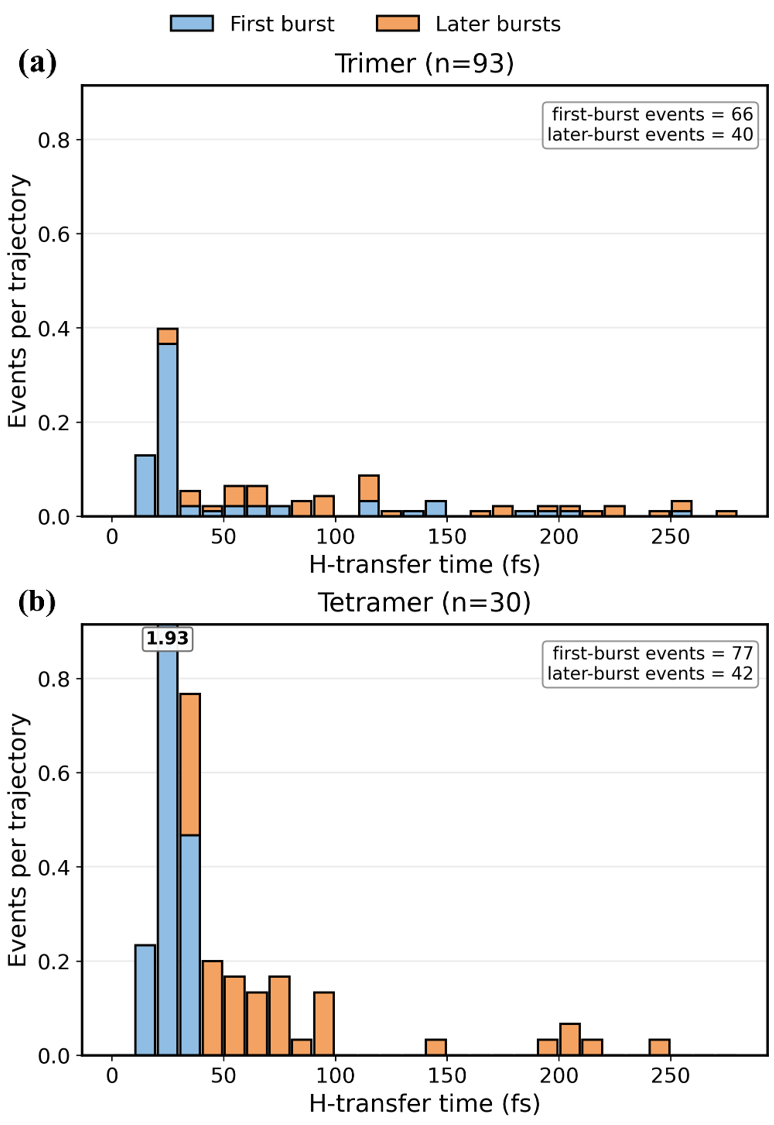}
  \caption{
Histograms showing the time distribution of H-transfer events for the trimer and tetramer
[panels (a)--(b), respectively]. Each bar corresponds to a 10-fs time interval, and the y-axis gives
the number of events per trajectory, calculated as the total number of H-transfer events occurring in
that interval divided by the total number of trajectories for the corresponding system.
The first burst is defined as the first H-transfer event in a trajectory together with all subsequent
H-transfer events occurring within 10~fs of that first event. All remaining H-transfer events are
classified as later-burst events. First-burst events are shown in blue, and later-burst events are shown
in orange.
}
  \label{transfer_time}
\end{figure}

FIG.~\ref{transfer_time}(a)--(b) show the time distributions of H-transfer events for the trimer
and tetramer, plotted as events per trajectory in 10-fs bins, using the same first-burst and
later-burst classification as in Sec.~\ref{sec:level5}. No timing histogram is shown for the
dimer, since only a single H-transfer event is detected across all dimer trajectories.

For the trimer [FIG.~\ref{transfer_time}(a)], the H-transfer signal is concentrated
predominantly in the early-time window. The 20--30~fs bin is the dominant feature, reaching
approximately 0.4 events per trajectory and consisting predominantly of first-burst events. Notably, small
contributions already appear in the 10--20~fs bin, which is earlier than the onset of H-ejection
in the trimer dataset ($\sim$20~fs). This early onset reflects the fact that H-transfer requires
only that the proton reach the acceptor oxygen --- a bond-stretching process --- whereas H-ejection
requires full O--H dissociation; transfer can therefore begin while the laser pulse is still
driving bond elongation. The first-burst-to-later-burst ratio across all trimer trajectories is
approximately 66:40 ($\sim$62:38), a noticeably larger later-burst contribution than was found
for H-ejection (70:30), indicating that secondary H-transfer --- proton migration triggered by
structural rearrangements after the initial pulse-driven event --- is a more prominent channel for
transfer than for ejection. Beyond the initial peak, the overall H-transfer intensity decreases
substantially but non-negligible contributions persist between 30 and 120~fs, dominated by
later-burst events and with all bins staying below $\sim$0.1 events per trajectory. After
120~fs only a weak tail remains. The trimer thus behaves as a crossover system in which the
pulse-driven early-transfer channel and a slower post-pulse structural rearrangement channel
both contribute appreciably to the total H-transfer yield.

For the tetramer [FIG.~\ref{transfer_time}(b)], the early-time peak is substantially enhanced
relative to the trimer. The 20--30~fs bin reaches 1.93 events per trajectory, nearly five times
the corresponding trimer peak of $\sim$0.4. This amplification is directly consistent with the
shorter initial O--O separations in the tetramer (2.73~\AA\ vs.\ 2.87--2.88~\AA\ in the trimer),
which lower the proton-migration barrier and place all four donor hydrogens within immediate
reach of their acceptor oxygens at the onset of the pulse. The first-burst-to-later-burst ratio
across all tetramer trajectories is approximately 77:42 ($\sim$65:35), similar to the trimer
ratio, confirming that secondary H-transfer contributes comparably in both systems despite the
much larger absolute early-time peak in the tetramer. Between 40 and 100~fs, later-burst events
remain clearly visible, reflecting sequential destabilization of further O--H bonds following the
initial pulse-driven cascade. Beyond 100~fs, the tetramer distribution shows a gradually
fading tail broadly similar to that of the trimer. The somewhat higher absolute activity in
this late-time region compared with the trimer is consistent with the larger number of
O--H bonds available for secondary rearrangement in the tetramer.

\textit{Comparison with H-ejection timing.}
A direct comparison between the H-ejection and H-transfer timing distributions shows a
systematic offset between the two processes. For both the trimer and tetramer, H-transfer begins slightly
earlier than H-ejection: transfer events appear in the 10--20~fs bin while ejection begins only
around 20~fs. This systematic offset reflects the lower energetic barrier for transfer relative
to ejection --- transfer requires only proton migration to a neighboring acceptor, while ejection
requires full O--H bond rupture and escape from the molecular framework. The coexistence of
both channels in the same time window during and immediately after the pulse indicates that
the strong-field ionization simultaneously activates two competing protonic pathways, with
geometry and network topology determining which prevails in a given trajectory.

The H-transfer statistics establish a clear and sharp size-dependent transition in the network-driven
protonic response. H-transfer is effectively absent in the dimer under the present strict criterion
and pulse conditions, reflecting the insufficiency of the brief ionization window and the single
hydrogen-bond topology to sustain stable proton migration. It becomes a majority channel in
the trimer (63\% per-trajectory probability), and reaches saturation in the tetramer (100\%),
making it the most discriminating observable for the dimer-to-trimer transition among all
quantities examined here. In the rate dimension, the dimer-to-trimer step produces an
approximately eightyfold increase in H-transfer events per monomer, far exceeding the
fourfold increase seen for H-ejection over the same step. In the timing dimension, both
the trimer and tetramer show strongly pulse-concurrent early-time peaks, but the later-burst
fraction is proportionally larger for H-transfer than for H-ejection in both systems, indicating
that secondary H-transfer --- driven by post-pulse structural relaxation of the ionized
network --- is a more persistent feature of the transfer dynamics than of ejection dynamics.

\subsection{\label{sec:level6E}Endpoint O--O Separation Statistics}

\begin{figure}[H]
  \centering
  \includegraphics[width=0.5\textwidth]{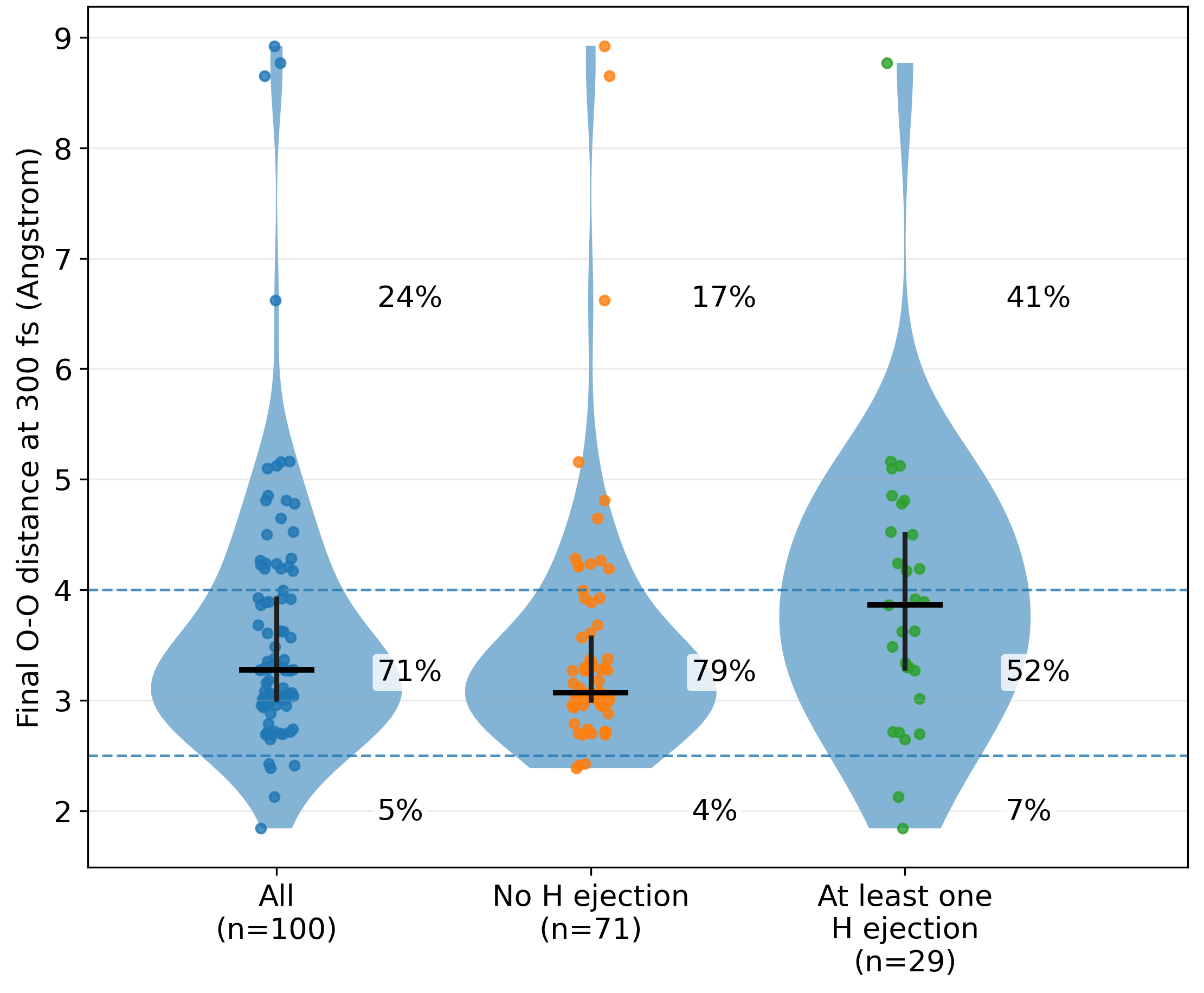}
  \caption{
Final O--O distance distributions for (\ce{H2O})$_2$ at the end of
the simulation ($t = 300$~fs), shown for all trajectories (left),
trajectories with no H-ejection (center), and trajectories
with at least one H-ejection (right).  Four trajectories in which an oxygen atom reaches the complex
absorbing potential (CAP) boundary are excluded. Individual points are jittered horizontally
for visibility; the horizontal line marks the median and the
vertical bar spans the interquartile range (Q1--Q3). Dashed lines mark empirical
thresholds: compression ($< 2.5$~\AA) and dissociation ($> 4.0$~\AA).
Percentages indicate the fraction of trajectories
in each region. H-ejection is strongly associated with dissociation: trajectories
without H-ejection predominantly remain near or below the initial
separation, while those with H-ejection are concentrated above the
dissociation threshold.
}
  \label{dimer_oo}
\end{figure}

\begin{figure}[H]
  \centering
  \includegraphics[width=0.5\textwidth]{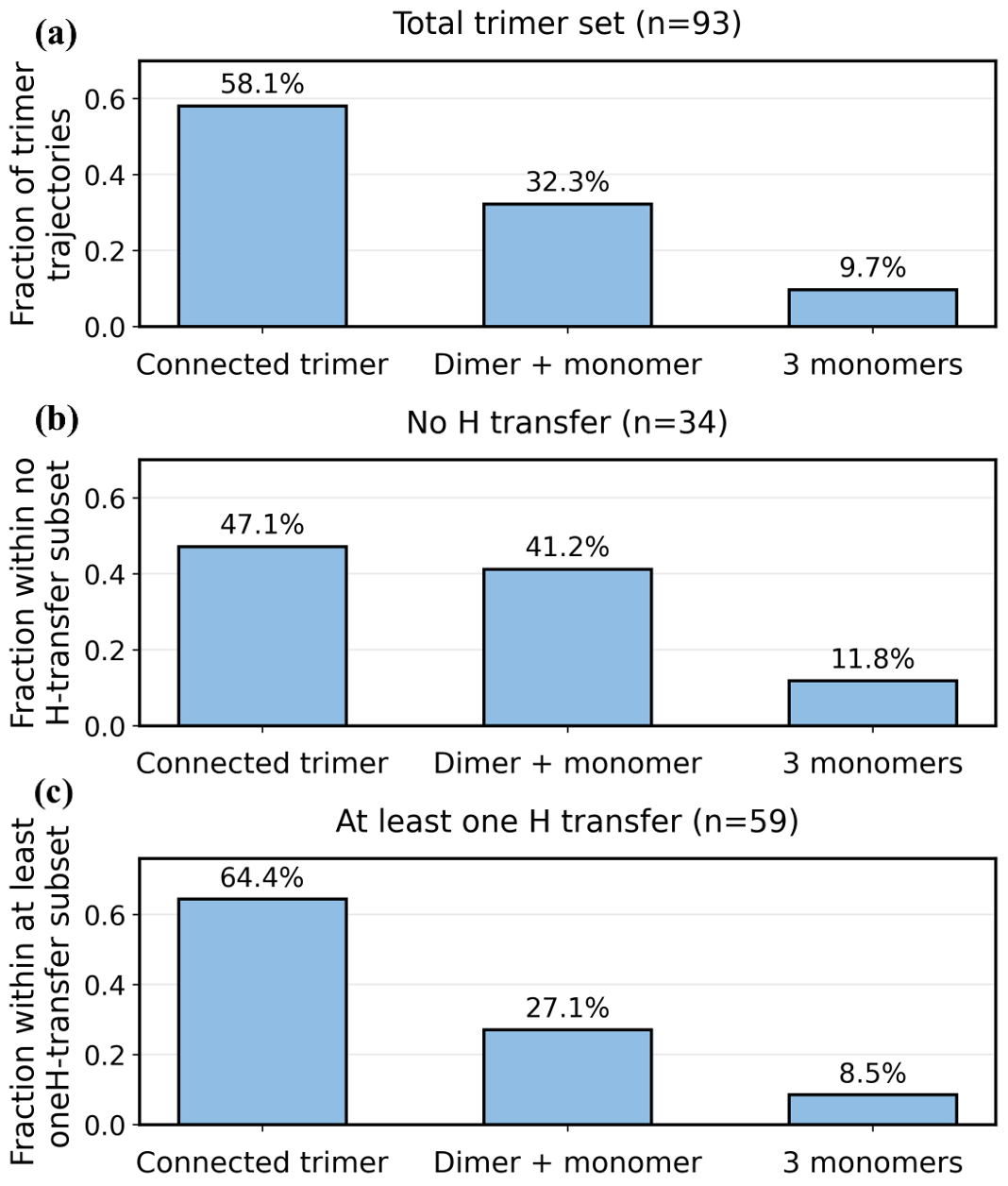}
  \caption{
Bar plots showing the percentages of the endpoint connectivity classes for the trimer:
(a) the full trajectory set, (b) the subset with no H-transfer, and (c) the subset with at least one
H-transfer. The three endpoint classes are connected trimer, dimer + monomer, and 3 monomers,
with the classification based on the 4.0~\AA\ dissociation threshold. The connected-trimer class
also includes rare open-chain cases in which one O--O pair exceeds 4.0~\AA, while the other two
O--O pairs remain below 4.0~\AA.
}
  \label{trimer_oo}
\end{figure}

\begin{figure}[H]
  \centering
  \includegraphics[width=0.5\textwidth]{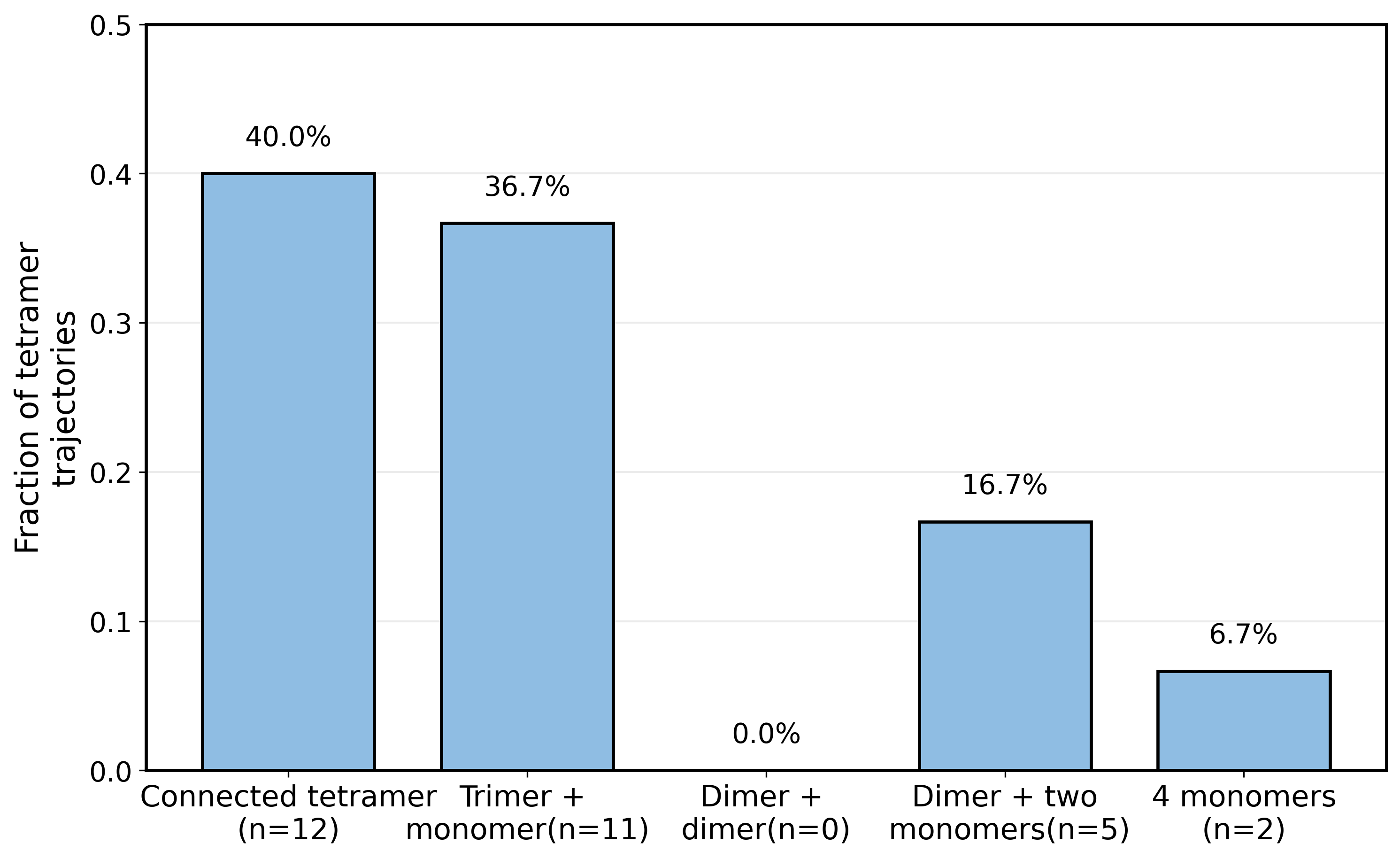}
  \caption{
Bar plots showing the percentages of the endpoint connectivity classes for the tetramer.
The five endpoint classes are connected tetramer, trimer + monomer, dimer + dimer,
dimer + two monomers, and 4 monomers, defined using the 4.0~\AA\ dissociation threshold.
The connected-tetramer class also includes rare open-chain cases in which one O--O pair exceeds
4.0~\AA, while the remaining oxygen framework still forms a single connected tetramer.
}
  \label{tetramer_oo}
\end{figure}

In this section, we present the statistics of the O--O separations at 300~fs for the dimer, trimer,
and tetramer. For endpoint classification, we introduce two operational thresholds: a compression
threshold of 2.5~\AA\ and a dissociation threshold of 4.0~\AA. These are not intended as strict
physical boundaries but as convenient reference values that divide the distribution into three
qualitatively distinct regimes --- strongly contracted, intermediate (near or modestly displaced from the initial separation, whether slightly compressed, unchanged, or slightly stretched), and dissociation-like --- within the present 300-fs observation window. Because the oxygen
framework evolves differently across cluster sizes, the dimer is characterized by the continuous
O--O distance distribution (FIG.~\ref{dimer_oo}), while the trimer and tetramer, which have
multiple O--O pairs, are characterized by discrete endpoint connectivity classes
(FIG.~\ref{trimer_oo} and FIG.~\ref{tetramer_oo}).

\textit{Dimer.}

FIG.~\ref{dimer_oo} shows the final O--O distance distribution for the water dimer. Four
trajectories are excluded because one oxygen atom reaches the CAP boundary during the
simulation; all four would correspond to very large O--O separations, and three of them contain
at least one H-ejection event while two contain double H-ejection. Including these four
CAP-reaching trajectories as dissociated-like cases, the total dimer dissociation probability
is $4/104 \times 100\% + 24\% \times (100/104) \approx 26.9\%$, a figure used in the
cross-size comparison below.

For the remaining 100 non-CAP trajectories, the dominant outcome across all three violin
subsets is the intermediate regime between 2.5 and 4.0~\AA, accounting for 71\%, 79\%, and
52\% of the all-trajectory, no-H-ejection, and at-least-one-H-ejection subsets, respectively.
The median O--O distance of the full distribution lies near 3.25~\AA, modestly above the
equilibrium dimer O--O separation of $\sim$2.91~\AA\ but below the dissociation threshold,
reflecting a typical endpoint in which the two oxygen-containing units are Coulomb-stretched
but not fully separated within the 300-fs window. Compressed endpoints below 2.5~\AA\ are
rare in all three groups (below 10\%), indicating that the net double ionization rarely drives the two oxygen centers closer together.

The most physically informative comparison is between the no-H-ejection and H-ejection subsets.
In the H-ejection subset, 41\% of trajectories reach the dissociation-like region above 4.0~\AA,
compared with only 17\% in the no-H-ejection subset --- a 2.4-fold difference. H-ejection is
therefore associated with larger O--O separations at the endpoint, consistent with the
expectation that proton loss reduces the number of bridging atoms between the two
oxygen-containing fragments. Notably, however, the 17\% dissociation-like fraction
among no-H-ejection trajectories shows that large O--O separation can also arise
from Coulomb repulsion between doubly ionized oxygen centers alone, without any proton loss
--- a distinct charge-separation channel that is present even when the O--H bonds remain
intact. The CAP-reaching subset reinforces this picture: three of the four excluded trajectories
contain H-ejection, consistent with the general trend that H-ejection strongly promotes
dissociation but is not its exclusive driver.

\textit{Trimer.}

For the trimer, a single O--O distance is no longer sufficient to characterize the endpoint
structure because three O--O pairs evolve simultaneously. The endpoint is therefore classified
into discrete connectivity classes based on the 4.0~\AA\ dissociation threshold applied to each
pair. The trimer set is separated according to H-transfer rather than H-ejection, because
H-ejection is already present in 90.3\% of trimer trajectories (Sec.~\ref{sec:level5}), making it
essentially universal and therefore uninformative as a separator; H-transfer, by contrast, occurs
in 63.44\% of trimer trajectories and thus still provides a meaningful binary partition.

For the full trimer ensemble [FIG.~\ref{trimer_oo}(a)], the connected-trimer class is the
largest single outcome (58.1\%), followed by the dimer + monomer channel (32.3\%), with
complete breakup into 3 monomers accounting for only 9.7\%. The overall dominance of connected
and partially connected endpoints confirms that the 300-fs window, while sufficient for extensive
protonic activity, is often too short for complete three-body oxygen-framework separation. If any
monomer separation beyond 4.0~\AA\ is counted as dissociation, the total trimer dissociation
probability is 42.0\% (32.3\% + 9.7\%), markedly higher than the dimer value of 26.9\% and
consistent with the trend of increasing oxygen breakup with cluster size.

The more revealing comparison comes from separating the trimer trajectories by H-transfer
[FIG.~\ref{trimer_oo}(b)--(c)]. In the no-transfer subset, the connected-trimer fraction
decreases to 47.1\% and the dimer + monomer channel rises to 41.2\%. In the subset with at
least one H-transfer event, the connected-trimer fraction increases substantially to 64.4\%
--- a 17 percentage-point swing relative to the no-transfer subset --- while the dimer + monomer
fraction drops to 27.1\% and the 3-monomer fraction remains small (8.5\%). H-transfer is therefore associated with a higher probability of preserving three-oxygen connectivity and suppressing partial breakup. The physical mechanism is electrostatic: when a proton migrates from one oxygen to a
neighboring one, it partially redistributes the net positive charge across the cluster, reducing the
inter-fragment charge imbalance and thereby weakening the Coulomb repulsion that would
otherwise drive the oxygens apart. Conversely, in trajectories without H-transfer the full charge
imbalance is concentrated on fewer bonds, producing stronger inter-fragment repulsion and a
higher probability of O--O dissociation.

\textit{Tetramer.}

For the tetramer, the endpoint classification expands to five connectivity classes reflecting the
richer fragmentation space of a four-oxygen system. Because every tetramer trajectory in the
present dataset exhibits at least one H-transfer event (Sec.~\ref{sec:level6D}), a no-transfer
subset cannot be formed and the tetramer results are shown as a single ensemble.

Among the five endpoint classes [FIG.~\ref{tetramer_oo}], the connected-tetramer and
trimer + monomer channels are essentially tied at 40.0\% and 36.7\%, respectively. This
near-tie is in stark contrast to the trimer, where the connected-trimer class leads the
dimer + monomer channel by almost 26 percentage points (58.1\% vs.\ 32.3\%). The remaining trajectories are distributed across dimer + two monomers (16.7\%) and
complete breakup into 4 monomers (6.7\%), while no dimer + dimer endpoint appears in the
present 30-trajectory dataset. The absence of the dimer + dimer channel in the present 30-trajectory dataset is noteworthy; while the limited sample size makes this observation tentative, it is also physically consistent with the cyclic-square topology of the tetramer ground-state structure: all O--O pairs are geometrically equivalent
in the ring, so asymmetric one-body detachment
(trimer + monomer) is kinematically more accessible than symmetric two-plus-two breakup,
which would require simultaneous
rupture of two opposing O--O pairs. Any conclusion regarding this channel should nonetheless remain tentative given the limited sample size.

If any monomer separation is counted as dissociation, the total tetramer dissociation
probability reaches 60.0\%, substantially higher than both the trimer (42.0\%) and
dimer (26.9\%) values and continuing the monotonic increase with cluster size. At the same
time, fully connected or only partially dissociated endpoints (connected tetramer +
trimer + monomer = 76.7\%) still account for more than three-quarters of all trajectories,
confirming that complete multi-body fragmentation within 300~fs remains uncommon even
in the tetramer.

The endpoint O--O statistics reveal a clear and systematic evolution of the oxygen-framework
response with increasing cluster size. In the dimer, H-ejection is associated with a higher
probability of large O--O expansion, while a secondary channel produces dissociation-like
endpoints even without proton loss. In the trimer and tetramer, where
multiple O--O pairs evolve simultaneously, the response is better described by connectivity
classes. H-transfer is associated with a higher probability of preserving connected oxygen
frameworks, consistent with proton redistribution partially reducing inter-fragment charge
imbalance. In the tetramer, the richer fragmentation landscape and the limited sample size (30
trajectories) make it difficult to draw strong conclusions about individual channels, though
the overall dissociation probability continues to rise.
Across all three systems, the dissociation probability increases monotonically with cluster size
--- 26.9\%, 42.0\%, and 60.0\% for the dimer, trimer, and tetramer, respectively --- reflecting
the combined effect of a larger number of O--O pairs available for rupture and the stronger
overall Coulomb repulsion in more highly charged larger clusters.

\subsection{\label{sec:KER}Comparison to Experiments}

The only system in the present dataset for which a clean experimental
data exists is the
water dimer, and within the dimer only the direct two-body Coulomb-explosion channel admits a
reliable comparison. We therefore focus this section on two aspects of the dimer comparison:
the kinetic energy release (KER) of the unprotonated channel, and the absence in our simulations
of the protonated two-body channel observed experimentally.

The reference experiment is the strong-field COLTRIMS study of Zhang~\textit{et al.}\cite{Zhang2019PRA},
who irradiated isolated water dimers with near-infrared laser pulses at $\lambda = 780$~nm,
a pulse duration of $38 \pm 2$~fs, and a peak intensity of
$(1.2 \pm 0.2)\times 10^{14}$~W/cm$^2$. They resolved two two-body Coulomb-explosion channels
following sequential double ionization within the same pulse: the unprotonated channel
(H$_2$O)$^+$ + (H$_2$O)$^+$ and the protonated channel (H$_3$O)$^+$ + OH$^+$. The
branching ratio between these two channels, $0.081 \pm 0.003$ in favor of the protonated
channel (approximately 1 protonated event per 12 double-ionization events), was used to infer
a proton-transfer time constant of $31 \pm 5$~fs for the singly charged dimer ion.

The present simulations use a central wavelength of 790~nm, a pulse duration of 6~fs (FWHM), and a peak field strength
of $E_{\max} = 7$~V/\AA, which corresponds to a peak intensity of approximately
$6.5\times10^{14}$~W/cm$^2$ --- roughly five times higher than the experimental value. The
much shorter pulse duration and higher peak intensity are therefore the two key differences
between the present conditions and the experiment, and both bear directly on the comparison
discussed below.

\textit{KER of the unprotonated channel.}

The most direct numerical comparison is the KER of the (H$_2$O)$^+$ + (H$_2$O)$^+$
channel. Of the 104 dimer trajectories, 13 (12.5\%) qualify as clean two-body breakup into
two intact H$_2$O-based fragments without H-ejection or CAP-reaching events, which is the
subset for which the asymptotic KER correction of Eq.~(12) can be applied reliably. Applying
the residual Coulomb-potential correction to these 13 trajectories, we obtain a mean asymptotic
KER of $4.47 \pm 1.03$~eV (mean $\pm$ standard deviation).

Zhang~\textit{et al.} reported a KER distribution center of 4.27~eV for the unprotonated
channel. Our mean of 4.47~eV lies only 0.20~eV (less than 5\%) above this value and well
within one-fifth of our standard deviation of 1.03~eV, indicating good quantitative agreement
given the statistical limitations of a 13-trajectory subset. The small remaining discrepancy is
further reduced when the shape of the experimental distribution is taken into account: the
reported distribution is slightly right-skewed, so its mean lies somewhat above the stated
distribution center of 4.27~eV, narrowing the gap between the experimental mean and our
computed value of 4.47~eV. The agreement therefore confirms that the RT-TDDFT/Ehrenfest
framework with the residual Coulomb correction of Eq.~(12) captures the energy scale of the
unprotonated two-body Coulomb-explosion channel to within the accuracy expected for a
13-trajectory comparison.

\textit{Absence of the protonated channel.}

\begin{figure}[H]
  \centering
  \includegraphics[width=0.5\textwidth]{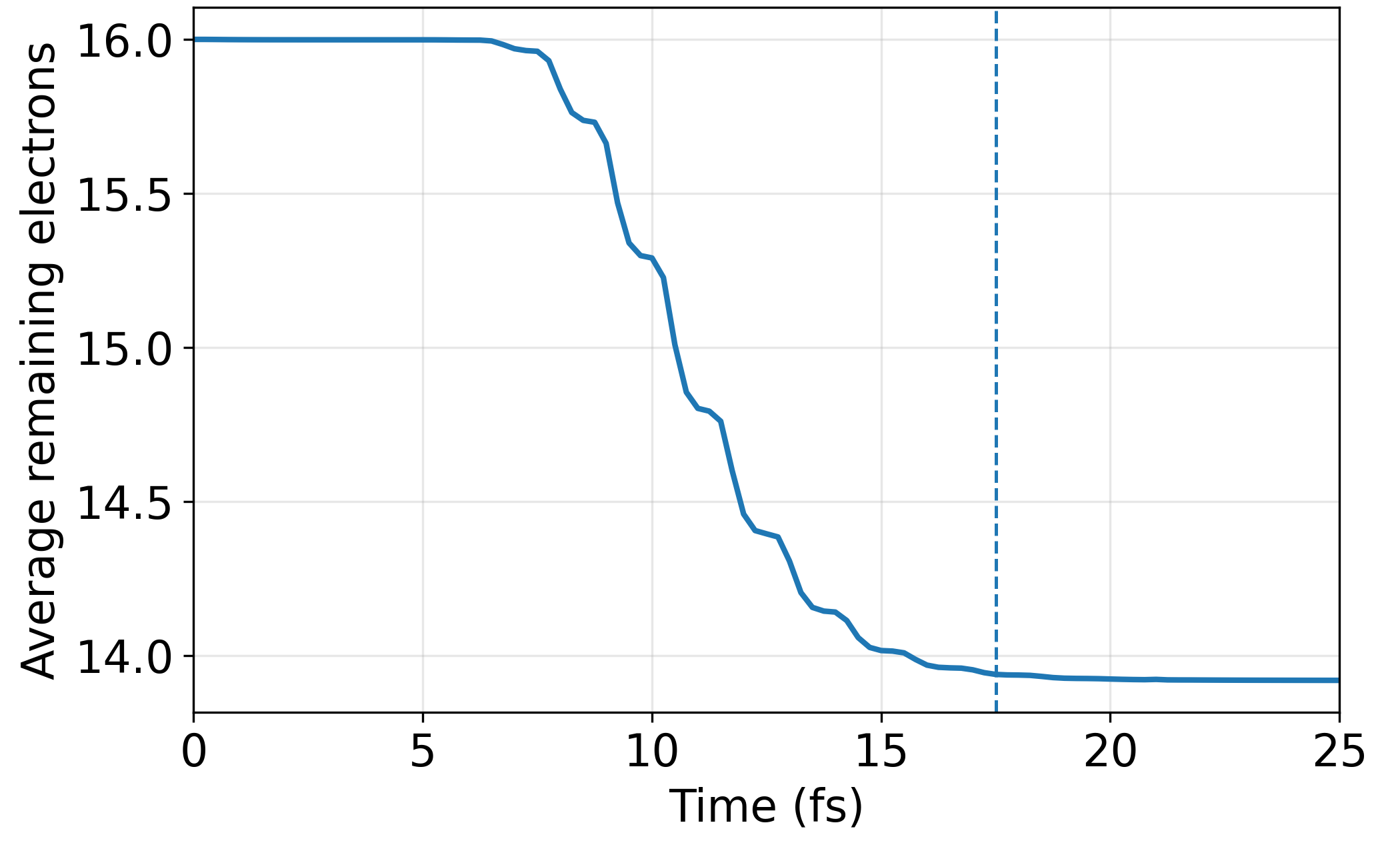}
  \caption{
Average remaining electrons for the full dimer trajectory ensemble as a function of time over the
first 25~fs of the simulation, corresponding to the pulse window. The data points are sampled at
0.25-fs intervals. The dotted vertical line marks 17.5~fs, the time by which the ionization has
reached 99\% of its final value at 25~fs.
}
  \label{e_time}
\end{figure}

We do not observe a clean protonated two-body Coulomb-explosion channel of the form
(H$_3$O)$^+$ + OH$^+$ in any of the 104 dimer trajectories. We do not regard this as a
fundamental contradiction with the experiment, but rather as a consequence of three distinct
differences between the present pulse conditions and those of Zhang~\textit{et al.}

First and most importantly, the effective ionization window in the present simulations is far
shorter than in the experiment. As shown in FIG.~\ref{e_time}, significant ionization begins
around 7~fs and 99\% of the final net ionization is already complete by $\sim$17.5~fs,
leaving an effective window of only $\sim$10~fs over which the ionization rises from near zero
to its final value. The
experimentally inferred proton-transfer time constant of $31 \pm 5$~fs is approximately
three times longer than this window. Under the present conditions, stable proton transfer is
therefore kinematically suppressed: the 10-fs ionization window is
insufficient for most trajectories to complete the $31$-fs transfer process before the
charge state reaches its final value and the opportunity for transfer is terminated.

Second, even for the small fraction of trajectories in which proton motion begins promptly
during the early stages of ionization, the strict 10-fs stability criterion imposed in Sec.~\ref{sec:level2}
(the transferred hydrogen must remain within 1.75~\AA\ of the new oxygen for 10 consecutive
femtoseconds) may not be satisfied before the charge buildup drives the system into a
higher charge configuration. The single H-transfer event detected in the dimer occurs at
134~fs, well after the ionization is complete, and represents a rare post-pulse rearrangement
rather than a pulse-concurrent transfer that would produce the protonated Coulomb-explosion
channel.

Third, the present peak intensity ($\sim6.5\times10^{14}$~W/cm$^2$) is approximately five
times higher than the experimental value ($(1.2\pm0.2)\times10^{14}$~W/cm$^2$). The higher
intensity accelerates the ionization process, further compressing the window available for
proton transfer and making the protonated
channel even less accessible under the present conditions.

Taken together, these three factors --- the $\sim$10-fs effective ionization window versus the
31-fs transfer timescale, the strict stability criterion, and the higher peak intensity --- offer
a consistent qualitative account for the absence of the protonated
two-body channel in the present simulations. The near-absence of stable dimer H-transfer
established in Sec.~\ref{sec:level6D} is in line with the same picture. Rather
than contradicting the experiment, the present results suggest that
the protonated channel is sensitive to pulse duration relative to the intrinsic proton-transfer timescale, and that the conditions required for its observation are not met in the present few-cycle,
high-intensity regime.

\section{Conclusion}

In this work, we have investigated the size-dependent strong-field ionization and dissociation
dynamics of (H$_2$O)$_n$ ($n=1$--4) under a common few-cycle near-infrared laser pulse using
real-time TDDFT coupled to Ehrenfest molecular dynamics. The central and overarching result is
that the size effect is far stronger in the protonic and oxygen-framework response than in the net
ionization itself: the mean ionization per monomer varies by less than 8\% across all four systems
(0.9823--1.0478~e per monomer), the pairwise differences in the means are small relative to the within-group spread, and the implied total charge states scale approximately as one electron
per monomer ($q \approx 1.0$, 2.1, 3.0, and 4.2 for the monomer through tetramer). These
results establish net charge deposition as a weakly size-dependent baseline, and confirm that the
dramatic differences described below arise from the hydrogen-bond-network topology and the
multi-center Coulomb interactions of a multiply ionized cluster rather than from differences in
total electron removal.

\textit{H-ejection.}
H-ejection is essentially absent in the isolated monomer (event rate 0.024 per monomer; 2.4\%
per-trajectory probability), establishing it as a network effect with no significant single-molecule
contribution. In the dimer, H-ejection is present but weak and temporally broad: the per-monomer
rate rises more than sixfold to 0.159 and the per-trajectory probability to 29.8\%, but first-burst events are
almost exclusively isolated single events (31 first-burst vs.\ 2 later-burst across all 104
trajectories) distributed broadly across the full 300-fs window, with some onset already at
20--30~fs during the trailing edge of the pulse. The dominant transition in H-ejection activity
occurs between the dimer and the trimer: the event rate increases more than fourfold (0.159
$\to$ 0.674~events per monomer) and the per-trajectory probability triples (29.8\%
$\to$ 90.3\%), while the response shifts from temporally broad and delayed to strongly
pulse-concurrent, with the 20--30~fs bin alone accounting for $\sim$0.8 events per trajectory in
the trimer and $\sim$1.83 in the tetramer. From trimer to tetramer, the total rate changes only
modestly (0.674 $\to$ 0.725), but the timing concentration intensifies further: the tetramer's
early-time peak is more than twice that of the trimer, and the signal from 30 to 300~fs is
dominated almost entirely by later-burst events. The overall H-ejection size dependence
therefore has two distinct dimensions: a large rate jump at the dimer--trimer boundary, and a
progressive concentration of timing into the earliest pulse window that continues to the tetramer.

\textit{H-transfer.}
H-transfer is the most sensitive indicator of the network-driven size transition. In the dimer,
stable H-transfer is effectively absent under the present strict criterion (0.0048 events per
monomer; 0.96\% per-trajectory probability), with the single detected event occurring at 134~fs
--- a rare post-pulse Coulomb-driven rearrangement rather than field-driven migration.
At the dimer--trimer boundary, the H-transfer event rate increases approximately eightyfold
(0.0048 $\to$ 0.38), and the per-trajectory probability rises from 0.96\% to 63.44\% --- a
transition that is nearly twenty times larger in magnitude than the corresponding H-ejection
step at the same boundary. In the tetramer, H-transfer saturates at a per-trajectory probability
of 100\% and a rate of 0.99 events per monomer, consistent with the cyclic-square topology
activating H-transfer across all four donor--acceptor pairs essentially without exception.
In the timing distributions, both the trimer and tetramer show strongly pulse-concurrent
early-time peaks, with H-transfer beginning slightly earlier than H-ejection (10--20~fs vs.\
$\sim$20~fs onset) because transfer requires only proton migration to a neighboring acceptor
rather than full O--H bond rupture. The later-burst fraction is proportionally larger for
H-transfer ($\sim$38\%) than for H-ejection ($\sim$30\%) in both systems, indicating that
secondary H-transfer driven by post-pulse structural relaxation is a more persistent
feature of transfer than of ejection dynamics.

\textit{Endpoint oxygen response.}
The endpoint oxygen-framework statistics reveal a clear evolution with cluster size.
In the dimer, H-ejection is associated with a higher probability of large O--O separation
(41\% of H-ejection trajectories reach the dissociation-like region above 4.0~\AA, vs.\ only
17\% without H-ejection), while
the 17\% dissociation-like fraction in the no-ejection subset identifies a secondary
channel in which doubly ionized oxygen centers repel without any proton loss.
In the trimer and tetramer, where multiple O--O pairs evolve simultaneously, the response is
better described by endpoint connectivity classes. H-transfer is associated with a higher
probability of preserving connected oxygen frameworks: trajectories with at least one
H-transfer event show a 17 percentage-point
higher connected-trimer fraction (64.4\% vs.\ 47.1\%) than those without. In the tetramer, the connected and trimer-plus-monomer
channels are essentially tied (40.0\% vs.\ 36.7\%), and the limited sample size (30 trajectories)
makes it difficult to draw strong conclusions about individual fragmentation channels.
Across all three systems, the dissociation probability increases monotonically ---
26.9\%, 42.0\%, and 60.0\% for the dimer, trimer, and tetramer --- indicating a systematic
trend toward greater oxygen-framework breakup with cluster size.

\textit{Comparison with experiment.}
The dimer two-body Coulomb-explosion KER obtained from the 13 clean direct-breakup
trajectories ($4.47 \pm 1.03$~eV) agrees with the experimental value of 4.27~eV reported
by Zhang~\textit{et al.}\cite{Zhang2019PRA} to within 5\%, and the remaining gap narrows
further when the right-skewed shape of the experimental distribution is taken into account.
The absence of the protonated (H$_3$O)$^+$ + OH$^+$ channel in our simulations is
accounted for by three compounding factors: the effective ionization window of $\sim$10~fs
(from onset at $\sim$7~fs to 99\% completion at $\sim$17.5~fs) is approximately three times
shorter than the experimentally inferred proton-transfer time constant of $31 \pm 5$~fs; the
strict 10-fs stability criterion is unlikely to be satisfied within this compressed window before
the charge state reaches its final value; and the present peak intensity
($\sim6.5\times10^{14}$~W/cm$^2$) is approximately five times higher than the experimental
value, further accelerating the ionization process. Rather than contradicting the experiment,
these results suggest that the protonated channel is sensitive to pulse duration relative to the
intrinsic proton-transfer timescale, and is suppressed under the
present few-cycle, high-intensity conditions.

\textit{Outlook.}
The present results demonstrate that increasing water-cluster size primarily reshapes the
strong-field response through proton-mediated and topology-level nuclear dynamics rather than
through changes in net ionization. Several directions for future work follow naturally.
First, orientation-resolved simulations would allow a direct test of the proposed structural
origin of the odd--even ionization pattern and of the orientation-dependent H-ejection
and H-transfer rates. Second, longer-pulse simulations at lower intensity --- conditions closer
to those of Zhang~\textit{et al.} --- would probe whether and at what pulse duration the
protonated two-body Coulomb-explosion channel re-emerges in the dimer, providing a
direct bridge between the present few-cycle regime and the nanosecond timescale of the
experimental proton-transfer dynamics. Third, extension to larger clusters ($n > 4$) would
test whether the H-transfer saturation observed at the tetramer persists for larger ring
structures, and whether new collective protonic pathways emerge in extended hydrogen-bond
networks. {\color{black}Together with the XUV pump--probe benchmark of Schnorr et al.~\cite{Schnorr2023SciAdv}, the present results provide a complementary picture spanning perturbative single-photon ionization and strong-field multi-ionization, showing how proton dynamics evolve from an isolated hydrogen bond to extended hydrogen-bond networks.} Finally, applying the same framework to other hydrogen-bonded systems --- ammonia
clusters, alcohol oligomers, or nucleobase pairs --- would determine whether the
topology-driven amplification of H-ejection and H-transfer observed here is a general feature
of hydrogen-bonded networks under strong-field multi-ionization, or specific to the
water-cluster geometry.

\begin{acknowledgments} 
This work was supported by the National
Science Foundation (NSF) 
under Grant No. DMR-2217759. Computational resources were provided by
ACES at 
Texas A$\&$M University through allocation PHYS240167 from the 
Advanced Cyberinfrastructure Coordination Ecosystem: Services $\&$
Support (ACCESS) program, 
supported by NSF grants 2138259, 2138286, 2138307, 2137603, and 2138296.
\end{acknowledgments}

\section*{Data Availability Statement}
All data and code are available at https://github.com/kvvandy/tddft.

\section*{AUTHOR DECLARATIONS}
\par\noindent
{\bf Conflict of Interest}

The authors have no conflict of interest to disclose.

%

\end{document}